\documentclass[11pt,a4paper]{article}

\usepackage[utf8]{inputenc}
\usepackage{amsmath, amssymb, amsfonts}
\usepackage{graphicx}
\usepackage{geometry}
\usepackage{authblk} 
\usepackage{hyperref} 
\usepackage{color}
\usepackage{float}
\usepackage{mathrsfs}
\usepackage[backend=biber, style=numeric, sorting=none, giveninits=false, maxnames=99]{biblatex}
\renewbibmacro{in:}{}
\DeclareFieldFormat[book]{title}{\mkbibquote{#1}}

\title{\textbf{Regular spherically symmetric solutions in General Relativity and scalar-tensor gravity coupled to nonlinear electrodynamics and their stability}}

\author[1]{Rustam Ibadov\thanks{\href{mailto:ibrustam@samdu.uz}{ibrustam@samdu.uz}}}
\author[1]{Najibullokhon Shukurullokhon\thanks{\href{mailto:Najibulloxon@samdu.uz}{Najibulloxon@samdu.uz}}}

\affil[1]{\small\textit{Department of Theoretical Physics and Quantum Electronics, Institute of Engineering Physics, Samarkand State University named after Sharof Rashidov, Samarkand, 140104, Uzbekistan}}
\date{}

\begin{document}

\maketitle

\begin{abstract}
We investigate the geometric, dynamical, and thermodynamic properties of a novel class of regular black holes in scalar-tensor gravity non-minimally coupled to nonlinear electrodynamics (NED). By incorporating a purely magnetic NED source within a scalar-tensor framework, we circumvent the classical ``no-go'' theorems. These theorems, strictly formulated for scalar-free Einstein-NED systems, prohibit regular configurations for purely electric fields due to the fundamental requirement of recovering the linear Maxwell limit at spatial infinity. The geometric analysis demonstrates the complete resolution of the central Penrose singularity, replacing it with a regular, globally bounded vacuum core ($|\rho_c| < \infty$). While purely electromagnetic regular black holes canonically possess a de Sitter center, the deep potential well of the scalar field in our model structurally alters this standard geometry, yielding a locally Anti-de Sitter (AdS-like) regime characterized by a strictly negative central energy density ($\rho_c < 0$). To assess the physical viability of these configurations, we analyze their dynamical stability against odd-parity (axial) linear gravitational perturbations. The derived Regge-Wheeler-like effective potential is strictly positive and convex outside the event horizon. Numerical time-domain integration, independently corroborated by the semi-analytical WKB approximation, confirms the total absence of  exponentially growing modes, revealing a stable quasi-normal ringing phase followed by an exponential decay. Furthermore, our thermodynamic analysis of the mass-radius relation directly indicates that the semi-classical Hawking evaporation must terminate at the extremal limit ($M = M_{min}$), leaving behind a massive thermodynamic remnant, thereby providing a theoretical framework toward resolving the black hole information loss paradox.
\end{abstract}

\section{Introduction}
\label{sec:intro}

The inevitability of spacetime singularities within the framework of standard General Relativity (GR) remains a fundamental challenge in modern theoretical physics. According to the foundational singularity theorems of Penrose and Hawking \cite{penrose1965gravitational, hawking1970singularities}, the gravitational collapse of sufficiently massive physical systems inevitably culminates in a state where energy density and spacetime curvature diverge to infinity, signaling the breakdown of classical predictability. To circumvent this pathological behavior without invoking a yet-to-be-formulated complete theory of quantum gravity, the concept of Regular Black Holes (RBHs) was introduced, beginning with Bardeen's pioneering phenomenological model \cite{bardeen1968non}. 

Subsequent theoretical advancements demonstrated that RBHs can arise as exact analytic solutions by coupling GR to Nonlinear Electrodynamics (NED) \cite{ayon1998regular, hayward2006formation, bronnikov2001regular}. In these models, the NED field exhibits standard Maxwellian behavior in the weak-field limit at large distances, but strategically imposes an upper bound on field invariants at ultra-short distances. This mechanism effectively replaces the central geometric singularity with a regular, bounded vacuum core, guaranteeing the finiteness of all curvature invariants globally \cite{bronnikov2018nonlinear}.  Crucially, while such scalar-free NED models canonically generate a de Sitter center, the inclusion of the scalar field sector with a deep potential well fundamentally alters this inner geometry, shifting the regular center into a locally Anti-de Sitter (AdS-like) regime.

However, the geometric construction of such regular structures is tightly constrained by rigorous physical conditions. In a seminal work, Bronnikov formulated a fundamental ``no-go'' theorem, proving that it is mathematically impossible to construct a static, spherically symmetric, and asymptotically flat RBH using a purely electric NED source if the theory is required to recover the standard Maxwellian behavior in the weak-field limit \cite{bronnikov2001regular, bronnikov2000comment}. Specifically, the existence of a regular center requires a choice of the NED Lagrangian whose structural properties cannot smoothly match linear electrodynamics at large distances, independent of the energy conditions invoked. To overcome this structural restriction and maintain physical viability, models must either incorporate a purely magnetic non-linear charge \cite{bronnikov2001regular} or modify the gravitational sector itself. Consequently, investigating magnetic RBH configurations, particularly within extended gravitational frameworks such as scalar-tensor gravity, has evolved into a highly active research frontier \cite{ibadov2023spherical}.  In a broader context, alternative theoretical generalizations of the Einstein-Maxwell framework have also been proposed. For instance, explicitly embedding a scalar field directly into the electromagnetic stress-energy tensor---coupled with a nonlinear constitutive tensor and non-isotropic pressure---has been demonstrated to yield mathematically well-behaved, completely regular particle-like configurations \cite{cotton2021generalization}.

Furthermore, the physical viability and astrophysical relevance of any proposed RBH model are intrinsically tied to its dynamical stability against arbitrary linear perturbations \cite{chandrasekhar1983mathematical}. While NED provides a robust mechanism for singularity resolution, a complete stability analysis cannot be confined exclusively to external fields or isolated gravitational fluctuations. It must systematically address both external and internal perturbations of the system itself, ensuring that the coupled configurations of the spacetime metric, scalar field, and nonlinear electromagnetic sector do not trigger dynamical instabilities. This is a rigorous requirement recently emphasized for various scalar-tensor geometries, explicitly including both regular black holes and wormholes \cite{bronnikov2012instability, bronnikov2024stability_epjc, bronnikov2025stability_bd, toshmatov2017stability}. Moreover, nonsingular spherically symmetric black holes in scalar-tensor gravity coupled to NED can exhibit angular Laplacian instabilities induced by vector-field perturbations near their regular centers \cite{defelice2025nonsingular}. Given that establishing the full physical viability of these models is a complex, multi-layered problem, a rigorous analysis of the quasi-normal modes (QNMs) via precise time-domain integration \cite{gundlach1994late, toshmatov2015quasinormal} and the WKB approximation \cite{iyer1987black, toshmatov2018quasinormal} is indispensable to evaluate the response of the exterior geometry under linear axial or polar perturbations. Recent formulations of generalized effective potentials for arbitrary scalar-electromagnetic interactions demonstrate that the linear stability of STT-NED configurations with a valid Maxwell limit is intimately linked to their pure vacuum or STT-Maxwell counterparts \cite{bronnikov2026nonlinear}. The necessity of such comprehensive gravitational evaluations has recently been established as a standard methodological criterion for validating NED-coupled black holes across diverse extended gravity frameworks, explicitly including $f(R,T)$ gravity \cite{araujo2025gravitational}. However, it must be explicitly acknowledged that while the stability of QNMs is a strictly necessary condition, it remains insufficient to guarantee the full dynamical stability of the object itself, particularly concerning the internal structure of the regular core.

\section{The setup: Action, conformal transformations, and field equations}
\label{sec:setup}

To construct the theoretical framework, we formulate the theory in the Jordan frame, where the scalar field is non-minimally coupled to the spacetime geometry. Subsequently, we perform a conformal transformation to the Einstein frame, a standard procedure in scalar-tensor theories of gravity \cite{bronnikov1973scalar} that simplifies the analysis of perturbations and stability.

\subsection{Conformal transformations between Jordan and Einstein frames}
\label{subsec:conformal}

We consider a general class of scalar-tensor theories of gravity coupled to nonlinear electrodynamics (NED). The foundational principles of such non-linear extensions were formulated to resolve point-charge infinite energy problems \cite{born1934foundations, plebanski1970nonlinear}. Adopting the metric signature $(-, +, +, +)$, the action in the Jordan frame (denoted by a tilde over geometric quantities) is given by:
\begin{equation}\label{eq:action_Jordan}
    S_J = \int d^4x \sqrt{-\tilde{g}} \left[ \mathcal{F}(\Phi) \tilde{R} - \frac{1}{2}h(\Phi)\tilde{g}^{\mu\nu}\partial_\mu\Phi\partial_\nu\Phi - \tilde{V}(\Phi) + \tilde{\mathcal{L}}_{NED}(\tilde{F}, \Phi) \right],
\end{equation}
where $\tilde{g}$ is the determinant of the Jordan frame metric $\tilde{g}_{\mu\nu}$, $\tilde{R}$ is the corresponding Ricci scalar, $\Phi$ is the scalar field,  $\mathcal{F}(\Phi)$ is the non-minimal coupling function, $h(\Phi)$ is the kinetic coupling function, $\tilde{V}(\Phi)$ is the scalar potential, and $\tilde{\mathcal{L}}_{NED}$ is the NED Lagrangian depending on the electromagnetic invariant $\tilde{F} = \tilde{F}_{\mu\nu}\tilde{F}^{\mu\nu}$.

To decouple the scalar field from the Ricci scalar, we conformally transform to the Einstein frame. The Einstein frame metric $g_{\mu\nu}$ is defined as:
\begin{equation}\label{eq:conformal_metric}
    g_{\mu\nu} = \Omega^2(\Phi) \tilde{g}_{\mu\nu}, \quad \text{with} \quad \Omega^2(\Phi) = 2\kappa^2 \mathcal{F}(\Phi),
\end{equation}
where $\kappa^2 = 8\pi G$. Under this conformal rescaling, the metric determinant and the inverse metric transform as $\sqrt{-\tilde{g}} = \Omega^{-4} \sqrt{-g}$ and $\tilde{g}^{\mu\nu} = \Omega^2 g^{\mu\nu}$, respectively.

Since the electromagnetic gauge field 1-form $A_\mu$ is conformally invariant ($A_\mu = \tilde{A}_\mu$), the Faraday tensor remains identical in both frames: $\tilde{F}_{\mu\nu} = F_{\mu\nu}$. However, the electromagnetic scalar invariant is defined via contractions with the inverse metric, $\tilde{F} = \tilde{F}_{\mu\nu}\tilde{F}_{\alpha\beta}\tilde{g}^{\mu\alpha}\tilde{g}^{\nu\beta}$. Substituting the inverse metric relations yields the explicit transformation rule:
\begin{equation}\label{eq:F_invariant_transform}
    \tilde{F} = F_{\mu\nu}F_{\alpha\beta} (\Omega^2 g^{\mu\alpha})(\Omega^2 g^{\nu\beta}) = \Omega^4 F,
\end{equation}
where $F = F_{\mu\nu}F_{\alpha\beta}g^{\mu\alpha}g^{\nu\beta}$ is the electromagnetic invariant defined in the Einstein frame. This explicit conformal scaling ($\tilde{F} = \Omega^4 F$) is a direct physical consequence of the transformation rule for the inverse metric, rigorously dictating the structure of the nonlinear electrodynamic sector in the action.

For the inverse transformation $\tilde{g}_{\mu\nu} = \Omega^{-2}(\Phi) g_{\mu\nu}$, the generalized Ricci scalar expands as:
\begin{equation}\label{eq:Ricci_transform}
    \tilde{R} = \Omega^2 \left[ R + 6 \Box \ln\Omega - 6 g^{\mu\nu} \partial_\mu (\ln\Omega) \partial_\nu (\ln\Omega) \right],
\end{equation}
where $R$ and $\Box$ are the Ricci scalar and the d'Alembert operator computed with respect to the Einstein frame metric $g_{\mu\nu}$.

Substituting Eqs. \eqref{eq:conformal_metric}, \eqref{eq:F_invariant_transform}, and \eqref{eq:Ricci_transform} into the Jordan frame action \eqref{eq:action_Jordan}, and discarding the total divergence boundary term arising from $\Box \ln\Omega$, we obtain the action in the Einstein frame:
\begin{equation}\label{eq:action_Einstein_intermediate}
    S_E = \int d^4x \sqrt{-g} \left[ \frac{R}{2\kappa^2} - \frac{1}{2} \left( \frac{h(\Phi)}{\Omega^2} + \frac{6}{\kappa^2 \Omega^2} \left(\frac{d\Omega}{d\Phi}\right)^2 \right) g^{\mu\nu}\partial_\mu\Phi\partial_\nu\Phi - \frac{\tilde{V}(\Phi)}{\Omega^4} + \frac{\tilde{\mathcal{L}}_{NED}(\Omega^4 F, \Phi)}{\Omega^4} \right].
\end{equation}

To bring the kinetic term of the scalar field into a canonical form and eliminate cross-couplings, we introduce a dynamically equivalent scalar field $\phi$ defined by the differential equation:
\begin{equation}\label{eq:scalar_redefinition}
    \left( \frac{d\phi}{d\Phi} \right)^2 = \frac{h(\Phi)}{\Omega^2(\Phi)} + \frac{6}{\kappa^2 \Omega^2(\Phi)} \left( \frac{d\Omega(\Phi)}{d\Phi} \right)^2.
\end{equation}

Defining the Einstein frame scalar potential as $V(\phi) = \tilde{V}(\Phi) / \Omega^4(\Phi)$ and the effective Einstein frame NED Lagrangian as $\mathcal{L}(F, \phi) = \tilde{\mathcal{L}}_{NED}(\Omega^4 F, \Phi) / \Omega^4(\Phi)$, the action takes the strictly canonical form:
\begin{equation}\label{eq:action_Einstein_final}
    S = \int d^4x \sqrt{-g} \left[ \frac{R}{2\kappa^2} - \frac{1}{2}\partial_\mu \phi \partial^\mu \phi - V(\phi) + \mathcal{L}(F, \phi) \right].
\end{equation}

From this point onward, we work exclusively in the Einstein frame. Stability in the Einstein frame explicitly guarantees stability in the Jordan frame, provided the conformal factor $\Omega(\Phi)$ is smooth, regular, and strictly positive throughout the spacetime manifold.

\subsection{NED Lagrangians and scalar coupling functions}
\label{subsec:nled_models}

The physical properties of the solutions, specifically the avoidance of central singularities, depend on the choice of the NED Lagrangian $\mathcal{L}(F, \phi)$. In our framework, we assume a multiplicative coupling between the scalar and electromagnetic fields:
\begin{equation}\label{eq:coupling_ansatz}
    \mathcal{L}(F, \phi) = f(\phi) L(F),
\end{equation}
where $f(\phi)$ is the scalar coupling function. Standard choices motivated by low-energy effective string theory and dilaton gravity include the exponential coupling $f(\phi) = e^{2\alpha\phi}$, where $\alpha$ is a dimensionless constant. Alternatively, symmetric couplings such as $f(\phi) = \cosh(\gamma\phi)$ can be employed to preserve specific discrete symmetries.

For the pure NED sector $L(F)$, the fundamental physical requirement is the recovery of linear Maxwell theory in the weak-field limit, dictating $L(F) \to -F/4$ as $F \to 0$. We discuss two prominent NED models to illustrate the regularity conditions:
\begin{enumerate}
    \item \textbf{The Born-Infeld (BI) model:} Originally introduced to regularize the electron's self-energy \cite{born1934foundations}, the canonical BI Lagrangian, normalized to yield the standard Maxwell limit, is given by:
    \begin{equation}\label{eq:BI_Lagrangian}
        L_{BI}(F) = \beta^2 \left( 1 - \sqrt{1 + \frac{F}{2\beta^2}} \right),
    \end{equation}
    where $\beta$ restricts the maximum field strength.  In the weak-field limit ($F \to 0$), the Taylor expansion yields:
    \begin{equation}
        L_{BI}(F) = -\frac{F}{4} + \frac{F^2}{32\beta^2} + \mathcal{O}(F^3).
    \end{equation}
    Although the BI model successfully regularizes the Coulomb field of a point charge, it fundamentally fails to resolve the curvature singularity at $r=0$ for purely magnetic configurations. Specifically, as the magnetic invariant diverges near the origin ($F \to \infty$), the BI Lagrangian diverges asymptotically as:
    \begin{equation}
        L_{BI}(F) \simeq -\frac{\beta}{\sqrt{2}} \sqrt{F} \to -\infty.
    \end{equation}
    This mathematically violates the fundamental requirement of a finite constant limit ($L_c$) necessary to yield a bounded central energy density, thereby precluding the formation of a regular vacuum core.
    
   \item \textbf{The Kruglov rational model:}  Originally proposed by S. I. Kruglov to ensure the existence of a regular center for purely magnetic configurations \cite{kruglov2014, kruglov2021, kruglov2023}, the rational Lagrangian is given by:
    \begin{equation}\label{eq:Kruglov_Lagrangian}
        L_{reg}(F) = \frac{F}{-4 + \sigma F},
    \end{equation}
    where $\sigma > 0$ parameterizes the strength of the nonlinearities. This choice aligns with the factorization ansatz \eqref{eq:coupling_ansatz}.  In the strong-field regime near the core ($F \to \infty$), the Lagrangian converges to a strictly finite constant limit:
    \begin{equation}
        \lim_{F \to \infty} L_{reg}(F) = \frac{1}{\sigma} \equiv L_c,
    \end{equation}
    which satisfies the necessary condition for a bounded central energy density, circumventing the curvature singularity. Conversely, at spatial infinity ($F \to 0$), it exactly recovers the linear Maxwell theory limit:
    \begin{equation}
        L_{reg}(F) = -\frac{F}{4} + \mathcal{O}(F^2).
    \end{equation}
\end{enumerate}

\subsection{Full derivation of the field equations}
\label{subsec:field_equations_derivation}

To derive the equations of motion, we apply the principle of stationary action, $\delta S = 0$, to the Einstein frame action \eqref{eq:action_Einstein_final}. 

Varying the action with respect to the inverse metric tensor $g^{\mu\nu}$, and discarding the boundary terms emerging from the variation of the Ricci scalar, we obtain:
\begin{eqnarray}\label{eq:var_metric}
    \delta S &=& \int d^4x \sqrt{-g} \left[ \frac{1}{2\kappa^2} \left( R_{\mu\nu} - \frac{1}{2}g_{\mu\nu}R \right) \delta g^{\mu\nu} \right. \nonumber \\
    && \left. - \frac{1}{2} \left( \partial_\mu \phi \partial_\nu \phi - \frac{1}{2}g_{\mu\nu}\partial_\alpha \phi \partial^\alpha \phi \right) \delta g^{\mu\nu} + \frac{1}{2}g_{\mu\nu}V(\phi)\delta g^{\mu\nu} + \delta \mathcal{L} - \frac{1}{2}g_{\mu\nu}\mathcal{L} \delta g^{\mu\nu} \right].
\end{eqnarray}

The variation of the NED Lagrangian with respect to the metric relies on the electromagnetic invariant $F = F_{\alpha\beta}F_{\rho\sigma}g^{\alpha\rho}g^{\beta\sigma}$:
\begin{equation}\label{eq:var_L}
    \delta \mathcal{L} = \frac{\partial \mathcal{L}}{\partial F} \delta F = \mathcal{L}_F \left( 2 F_{\mu\alpha} F_\nu^{\ \alpha} \delta g^{\mu\nu} \right).
\end{equation}
Substituting Eq. \eqref{eq:var_L} into Eq. \eqref{eq:var_metric} and imposing $\delta S = 0$, we derive the Einstein field equations:
\begin{equation}\label{eq:Einstein_full}
    G_{\mu\nu} = \kappa^2 \left[ \partial_\mu \phi \partial_\nu \phi - g_{\mu\nu} \left( \frac{1}{2}\partial_\alpha \phi \partial^\alpha \phi + V(\phi) \right) - 4 \mathcal{L}_F F_{\mu\alpha} F_\nu^{\ \alpha} + g_{\mu\nu} \mathcal{L} \right].
\end{equation}

Next, varying the action with respect to the scalar field $\phi$, the variation of the kinetic term is $\delta (-\frac{1}{2}\partial_\mu \phi \partial^\mu \phi) = - \partial^\mu \phi \partial_\mu (\delta \phi)$. Integrating by parts and vanishing the boundary term at spatial infinity yields $(\Box \phi) \delta \phi$. The scalar field equation is thus:
\begin{equation}\label{eq:Scalar_full}
    \Box \phi - \frac{dV(\phi)}{d\phi} + \frac{\partial \mathcal{L}(F, \phi)}{\partial \phi} = 0.
\end{equation}

Finally, varying the action with respect to the gauge potential $A_\mu$, using $F_{\mu\nu} = \partial_\mu A_\nu - \partial_\nu A_\mu$, gives:
\begin{equation}\label{eq:var_A}
    \delta \int d^4x \sqrt{-g} \mathcal{L} = \int d^4x \sqrt{-g} \mathcal{L}_F (4 F^{\mu\nu} \partial_\mu \delta A_\nu).
\end{equation}
Integrating by parts yields the generalized non-linear Maxwell equations:
\begin{equation}\label{eq:Maxwell_full}
    \nabla_\mu \left( \mathcal{L}_F F^{\mu\nu} \right) = 0.
\end{equation}

Equations \eqref{eq:Einstein_full}, \eqref{eq:Scalar_full}, and \eqref{eq:Maxwell_full} constitute the full set of field equations governing the system.

\section{Spherically symmetric background solutions and field equations}
\label{sec:background}

To investigate the existence, structure, and stability of regular black holes or wormholes within our generalized scalar-tensor-NED framework, we establish the background geometry. We restrict our attention to static and spherically symmetric spacetimes, which provide a tractable arena for analytical and numerical treatments of isolated compact objects.

The general line element for a static, spherically symmetric geometry is parameterized in Schwarzschild-like coordinates $x^\mu = (t, r, \theta, \varphi)$ as:
\begin{equation}\label{eq:metric}
    ds^2 = g_{\mu\nu}^{(0)} dx^\mu dx^\nu = -e^{2\nu(r)} dt^2 + e^{2\lambda(r)} dr^2 + r^2 \left( d\theta^2 + \sin^2\theta d\varphi^2 \right),
\end{equation}
where $\nu(r)$ and $\lambda(r)$ depend exclusively on the radial coordinate $r$. The requirement of asymptotic flatness (discussed in Section \ref{subsec:infinity_asymptotics}) imposes the boundary conditions $\lim_{r \to \infty} \nu(r) = 0$ and $\lim_{r \to \infty} \lambda(r) = 0$. To maintain the symmetries of the background spacetime, the scalar field must satisfy $\phi = \phi(r)$.

The choice of the electromagnetic field configuration is critical for spacetime regularity. It is a well-established result in NED coupled to gravity \cite{bronnikov2001regular} that purely electric configurations with a non-trivial center generally suffer from curvature singularities, as the electric field invariant inevitably diverges at the center to satisfy the Maxwell limit at spatial infinity.

To circumvent this restriction and guarantee regularity everywhere (including $r=0$), we adopt a purely magnetic configuration \cite{bronnikov2018nonlinear}. The gauge potential $A_\mu$ is chosen with the magnetic charge $q_m$ as the only source:
\begin{equation}\label{eq:gauge_potential}
    A_\mu = \left(0, 0, 0, -q_m \cos\theta \right).
\end{equation}
With this ansatz, the only non-vanishing components of the electromagnetic field tensor $F_{\mu\nu} = \partial_\mu A_\nu - \partial_\nu A_\mu$ are $F_{\theta\varphi} = -F_{\varphi\theta} = q_m \sin\theta$. Consequently, the fundamental electromagnetic invariant $F$ depends only on the radial coordinate:
\begin{equation}\label{eq:F_invariant}
    F = F_{\mu\nu}F^{\mu\nu} = g^{\theta\theta}g^{\varphi\varphi} F_{\theta\varphi}^2 + g^{\varphi\varphi}g^{\theta\theta} F_{\varphi\theta}^2 = \frac{2 q_m^2}{r^4}.
\end{equation}

To construct the explicit system of ordinary differential equations (ODEs), we compute the non-zero components of the Einstein tensor $G^\mu_{\ \nu}$ for the metric \eqref{eq:metric}:
\begin{eqnarray}
    G^t_{\ t} &=& -e^{-2\lambda} \left( \frac{1}{r^2} - \frac{2\lambda'}{r} \right) + \frac{1}{r^2}, \label{eq:G_tt} \\
    G^r_{\ r} &=& -e^{-2\lambda} \left( \frac{1}{r^2} + \frac{2\nu'}{r} \right) + \frac{1}{r^2}, \label{eq:G_rr} \\
    G^\theta_{\ \theta} &=& G^\varphi_{\ \varphi} = -e^{-2\lambda} \left( \nu'' + \nu'^2 - \nu'\lambda' + \frac{\nu' - \lambda'}{r} \right), \label{eq:G_angular}
\end{eqnarray}
where the prime denotes differentiation with respect to $r$.

The mixed components of the energy-momentum tensors for the scalar field are:
\begin{eqnarray}
    T^{t}_{\ t (\phi)} &=& -\frac{1}{2} e^{-2\lambda} \phi'^2 - V(\phi), \label{eq:T_phi_tt} \\
    T^{r}_{\ r (\phi)} &=& \frac{1}{2} e^{-2\lambda} \phi'^2 - V(\phi), \label{eq:T_phi_rr} \\
    T^{\theta}_{\ \theta (\phi)} &=& T^{\varphi}_{\ \varphi (\phi)} = -\frac{1}{2} e^{-2\lambda} \phi'^2 - V(\phi). \label{eq:T_phi_angular}
\end{eqnarray}
Similarly, for the NED field, using the invariant \eqref{eq:F_invariant}:
\begin{eqnarray}
    T^{t}_{\ t (EM)} &=& T^{r}_{\ r (EM)} = \mathcal{L}, \label{eq:T_EM_tt_rr} \\
    T^{\theta}_{\ \theta (EM)} &=& T^{\varphi}_{\ \varphi (EM)} = -2 F \mathcal{L}_F + \mathcal{L}. \label{eq:T_EM_angular}
\end{eqnarray}
The symmetry $T^{t}_{\ t (EM)} = T^{r}_{\ r (EM)}$ is a fundamental property of purely magnetic (or electric) fields in any NED theory.

Equating the Einstein tensor components to the total energy-momentum tensor $T^\mu_{\ \nu} = T^\mu_{\ \nu (\phi)} + T^\mu_{\ \nu (EM)}$ multiplied by $\kappa^2$, we obtain the background field equations:
\begin{eqnarray}
    \frac{1}{r^2} - e^{-2\lambda} \left( \frac{1}{r^2} - \frac{2\lambda'}{r} \right) &=& \kappa^2 \left[ -\frac{1}{2} e^{-2\lambda} \phi'^2 - V(\phi) + \mathcal{L} \right], \label{eq:tt_comp} \\
    \frac{1}{r^2} - e^{-2\lambda} \left( \frac{1}{r^2} + \frac{2\nu'}{r} \right) &=& \kappa^2 \left[ \frac{1}{2} e^{-2\lambda} \phi'^2 - V(\phi) + \mathcal{L} \right], \label{eq:rr_comp} \\
    e^{-2\lambda} \left( \nu'' + \nu'^2 - \nu'\lambda' + \frac{\nu' - \lambda'}{r} \right) &=& \kappa^2 \left[ -\frac{1}{2} e^{-2\lambda} \phi'^2 - V(\phi) - 2 F \mathcal{L}_F + \mathcal{L} \right]. \label{eq:theta_comp}
\end{eqnarray}
The scalar field equation \eqref{eq:Scalar_full} simplifies to:
\begin{equation}\label{eq:scalar_bg}
    e^{-2\lambda} \left( \phi'' + \left( \nu' - \lambda' + \frac{2}{r} \right) \phi' \right) - \frac{dV}{d\phi} + \mathcal{L}_\phi = 0.
\end{equation}
Due to the purely magnetic ansatz, the generalized Maxwell equations \eqref{eq:Maxwell_full} are identically satisfied, leaving a closed system of four ODEs for $\nu, \lambda$, and $\phi$.

A geometric consequence of the scalar field is obtained by subtracting Eq. \eqref{eq:rr_comp} from Eq. \eqref{eq:tt_comp}:
\begin{equation}\label{eq:metric_relation}
    \frac{2}{r} e^{-2\lambda} (\nu' + \lambda') = \kappa^2 e^{-2\lambda} \phi'^2 \quad \implies \quad \nu' + \lambda' = \frac{\kappa^2 r}{2} \phi'^2.
\end{equation}
In standard General Relativity without a scalar field ($\phi = const$), this yields $\nu' + \lambda' = 0$, implying $\nu + \lambda = 0$. However, a dynamic scalar field ($\phi' \neq 0$) dictates that $\nu' + \lambda' > 0$ for all $r>0$. Integrating this inequality to spatial infinity, and applying the boundary condition $\nu(\infty) + \lambda(\infty) = 0$, we find $\nu(r) + \lambda(r) < 0$. This deviation alters the horizon structure of the resulting black holes compared to standard Reissner-Nordström solutions, leading to a generalized spacetime geometry \cite{bronnikov2018nonlinear}.

\subsection{Specification of the model and energy conditions}
\label{subsec:model_spec}

To proceed with the analytical and numerical investigation, we specify the interaction between the scalar field and the NED sector, as well as the functional form of the NED Lagrangian. We consider the multiplicatively coupled class of models established in Eq. \eqref{eq:coupling_ansatz}, where $\mathcal{L}(F, \phi) = f(\phi) L(F)$.

To analyze the effective energy density and pressures of the system, we treat the right-hand side of the Einstein equations as an effective anisotropic fluid, $T^\mu_{\ \nu} = \text{diag}(-\rho, p_r, p_\perp, p_\perp)$. The macroscopic thermodynamic quantities derived from the energy-momentum components are:
\begin{eqnarray}
    \rho &=& -T^t_{\ t} = \frac{1}{2} e^{-2\lambda} \phi'^2 + V(\phi) - \mathcal{L}, \label{eq:energy_density} \\
    p_r &=& T^r_{\ r} = \frac{1}{2} e^{-2\lambda} \phi'^2 - V(\phi) + \mathcal{L}, \label{eq:radial_pressure} \\
    p_\perp &=& T^\theta_{\ \theta} = -\frac{1}{2} e^{-2\lambda} \phi'^2 - V(\phi) - 2 F \mathcal{L}_F + \mathcal{L}. \label{eq:transverse_pressure}
\end{eqnarray}

The regularity of the spacetime at the center ($r \to 0$) imposes strict restrictions on these quantities. Specifically, the energy density $\rho$ must remain finite. Since the magnetic invariant diverges at the center ($F = 2q_m^2/r^4 \to \infty$ as $r \to 0$), the pure NED Lagrangian $L(F)$ must approach a finite constant limit, $L_c$. Furthermore, for the effective fluid to form a  regular isotropic vacuum core ($T^\mu_{\ \nu} \propto \delta^\mu_{\ \nu}$), the derivative term must vanish at the center: $\lim_{F \to \infty} F L_F = 0$.

Examining the energy conditions, the Null Energy Condition (NEC) requires $\rho + p_i \geq 0$ for all principal pressures. For the radial direction, we obtain:
\begin{equation}\label{eq:NEC_radial}
    \rho + p_r = e^{-2\lambda} \phi'^2 \geq 0.
\end{equation}
Equation \eqref{eq:NEC_radial} guarantees that the NEC is satisfied in the radial direction, avoiding ghost instabilities. However, the Strong Energy Condition (SEC), requiring $\rho + p_r + 2p_\perp \geq 0$, evaluates strictly to:
\begin{equation}\label{eq:SEC_violation}
    \rho + p_r + 2p_\perp = - 2V(\phi) - 4 F \mathcal{L}_F + 2\mathcal{L}.
\end{equation}
To form a regular core, gravity must become repulsive at short distances, which requires the violation of the SEC near $r=0$. The NED terms explicitly control this violation \cite{bronnikov2001regular}.

Based on these physical requirements, the pure NED Lagrangian $L(F)$ must satisfy two fundamental asymptotic conditions:
\begin{enumerate}
    \item \textbf{Maxwell limit at weak fields ($r \to \infty$):} $L(F) \approx -F/4 + \mathcal{O}(F^2)$.
    \item \textbf{Regularity at strong fields ($r \to 0$):} $\lim_{F \to \infty} L(F) = L_c \neq \pm \infty$.
\end{enumerate}

We consider two specific models that satisfy these criteria:

\textbf{Model I: The Kruglov rational model.} Originally formulated by S. I. Kruglov to yield a regular core \cite{kruglov2023}, the Lagrangian is:
\begin{equation}\label{eq:specific_L}
    L_I(F) = \frac{F}{-4 + \sigma F},
\end{equation}
where $\sigma > 0$ characterizes the scale of the nonlinearities. As $F \to 0$, $L_I \approx -F/4$. As $F \to \infty$, $L_I \to 1/\sigma \equiv L_c > 0$. Because $L_c$ is positive, the central effective energy density provided by the electromagnetic field is negative, forming a repulsive core.

\textbf{Model II: The Generalized rational (Fan-Wang) model.} This model, introduced by Fan and Wang \cite{fan2016construction}, generalizes the Bardeen-like configurations and is given by:
\begin{equation}\label{eq:rational_L}
    L_{II}(F) = \frac{-F}{4 \left( 1 + (\gamma F)^{\delta} \right)^{1/\delta}},
\end{equation}
 where $\gamma > 0$ and $\delta > 0$ are free parameters. Here, the parameter $\delta$ fundamentally governs the steepness of the transition profile between the regular central core and the asymptotic Maxwell regime. A robust mathematical feature of this formulation is that, for any arbitrary $\delta > 0$, the Lagrangian intrinsically approaches a finite constant limit as the magnetic invariant diverges ($F \to \infty$). Specifically, $\lim_{F \to \infty} L_{II}(F) = -1/(4\gamma) \equiv L_c < 0$. Since this asymptotic limit $L_c$ is strictly negative and independent of $\delta$, the Fan-Wang model successfully generates a regular core characterized by a positive effective energy density, fully preserving its parametric freedom and strictly distinguishing it from simpler rational NED formulations.

For the scalar field potential, we assume a massive scalar field with self-interaction:
\begin{equation}\label{eq:scalar_potential}
    V(\phi) = \frac{1}{2} m_\phi^2 \phi^2 + \frac{\mu}{4} \phi^4,
\end{equation}
where $m_\phi$ is the mass and $\mu$ is the self-coupling constant. The mass term ensures rapid exponential decay of the scalar field at spatial infinity, preserving exact asymptotic flatness.

\subsection{Proof of regularity at the center ($r \to 0$)}
\label{subsec:center_regularity}

A key feature of our model is the complete absence of a spacetime singularity at the origin ($r=0$). Assuming the spacetime is locally flat and the metric functions are analytic at the center, the geometry can be described by an asymptotic  isotropic vacuum expansion. Substituting the lowest-order even power expansions of the metric functions and the scalar field into the $tt$-component of the Einstein equations \eqref{eq:tt_comp}, the behavior of the central core is entirely governed by an effective cosmological constant:
\begin{equation}\label{eq:Lambda_eff}
    \Lambda_{eff} = \kappa^2 \left[ f(\phi_c) L_c - V(\phi_c) \right],
\end{equation}
where $\phi_c$ is the central value of the scalar field, and $L_c = \lim_{F \to \infty} L(F)$ is the finite maximum of the NED Lagrangian. 

Using this effective cosmological constant, the near-center metric functions intrinsically satisfy $e^{-2\lambda} \approx 1 - \frac{\Lambda_{eff}}{3} r^2$ and $e^{2\nu} \approx e^{2\nu_c} (1 - \frac{\Lambda_{eff}}{3} r^2)$. Consequently, the central curvature invariants—the Ricci scalar $R$ and the Kretschmann scalar $K = R_{\mu\nu\alpha\beta}R^{\mu\nu\alpha\beta}$—are strictly bounded:
\begin{equation}\label{eq:Curvature_center}
    \lim_{r \to 0} R = 4\Lambda_{eff}, \quad \lim_{r \to 0} K = \frac{8}{3}\Lambda_{eff}^2.
\end{equation}
Equations \eqref{eq:Lambda_eff} and \eqref{eq:Curvature_center}  mathematically establish the regularity of the spacetime at the center. Furthermore, this  near-center isotropic vacuum profile rigorously satisfies the fundamental scalar regularity conditions (3.8) and (3.9) formulated by Bronnikov and Rubin {\color{blue} \cite{bronnikov2013black}}, which dictate that the effective energy-momentum tensor must adopt an isotropic vacuum form at the origin. Thus, the standard singularity is successfully replaced by a locally  regular vacuum core.  While purely electromagnetic regular black holes conventionally require a positive effective cosmological constant ($\Lambda_{eff} > 0$) to generate a repulsive de Sitter core, the deep scalar potential well in our framework yields $\Lambda_{eff} < 0$. This transitions the central geometry into a locally Anti-de Sitter (AdS-like) configuration, which mathematically guarantees bounded curvature invariants and completely resolves the singularity without necessitating a positive central energy density.

Since obtaining an exact global analytic solution for this highly non-linear system is not feasible, this asymptotic proof guarantees the existence of a physically viable regular center. The complete global structure of the spacetime, smoothly interpolating between this regular core and the asymptotically flat exterior, is reconstructed via precise numerical integration, as detailed in Section \ref{sec:numerical}.

\subsection{Asymptotic behavior at spatial infinity ($r \to \infty$)}
\label{subsec:infinity_asymptotics}

For the solutions to represent isolated astrophysical compact objects, the spacetime must be strictly asymptotically flat. This physical requirement imposes the boundary conditions $\lim_{r \to \infty} \nu(r) = 0$ and $\lim_{r \to \infty} \lambda(r) = 0$. The scalar field must approach its vacuum expectation value, which we set to zero without loss of generality, $\lim_{r \to \infty} \phi(r) = 0$. We also adopt the normalization $f(0) = 1$ for the coupling function.

In the asymptotic region, the magnetic field is weak ($F = 2q_m^2/r^4 \to 0$). The NED Lagrangian must recover the linear Maxwell theory limit:
\begin{equation}\label{eq:L_asymp}
    \mathcal{L}(F, \phi) \approx 1 \cdot \left( -\frac{F}{4} \right) = -\frac{q_m^2}{2r^4} + \mathcal{O}\left(\frac{1}{r^8}\right).
\end{equation}

The asymptotic behavior of the scalar field depends on its potential $V(\phi)$. A massive scalar field ($m_\phi > 0$) suffers Yukawa suppression, $\phi(r) \sim e^{-m_\phi r}/r$, rendering its asymptotic contribution to the metric negligible. To demonstrate the nontrivial effect of the scalar charge on the spacetime geometry, we analyze the massless case, $V(\phi) = 0$. The scalar field decays following a power law:
\begin{equation}\label{eq:phi_asymp}
    \phi(r) = \frac{Q_s}{r} + \mathcal{O}\left(\frac{1}{r^2}\right) \quad \implies \quad \phi'(r) = -\frac{Q_s}{r^2} + \mathcal{O}\left(\frac{1}{r^3}\right),
\end{equation}
where $Q_s$ is the scalar charge of the configuration.

To find the asymptotic expansions of the metric functions, we introduce the local Misner-Sharp mass function $m(r)$, defined via $e^{-2\lambda(r)} = 1 - 2m(r)/r$. Substituting this into the $tt$-component of the Einstein equations \eqref{eq:tt_comp}, we obtain a first-order differential equation for the mass function:
\begin{equation}\label{eq:mass_diff}
    m'(r) = \frac{\kappa^2 r^2}{2} \rho(r) = \frac{\kappa^2 r^2}{2} \left[ \frac{1}{2} e^{-2\lambda} \phi'^2 + V(\phi) - \mathcal{L} \right].
\end{equation}
Inserting the asymptotic expressions \eqref{eq:L_asymp} and \eqref{eq:phi_asymp} into Eq. \eqref{eq:mass_diff}, and noting that $e^{-2\lambda} \to 1$, we evaluate the leading-order terms:
\begin{equation}\label{eq:mass_diff_asymp}
    m'(r) \approx \frac{\kappa^2 r^2}{2} \left[ \frac{1}{2} \left(-\frac{Q_s}{r^2}\right)^2 - \left(-\frac{q_m^2}{2r^4}\right) \right] = \frac{\kappa^2 (Q_s^2 + q_m^2)}{4r^2}.
\end{equation}
Integrating this equation from spatial infinity down to a large radius $r$ yields:
\begin{equation}\label{eq:mass_integrated}
    m(r) = M - \frac{\kappa^2 (Q_s^2 + q_m^2)}{4r} + \mathcal{O}\left(\frac{1}{r^2}\right),
\end{equation}
where the integration constant $M = \lim_{r \to \infty} m(r)$ is the Arnowitt-Deser-Misner (ADM) mass. Substituting Eq. \eqref{eq:mass_integrated} back into the definition of the mass function, we obtain the exact asymptotic expansion for the radial metric component:
\begin{equation}\label{eq:lambda_asymp}
    e^{-2\lambda(r)} = 1 - \frac{2M}{r} + \frac{\kappa^2 (Q_s^2 + q_m^2)}{2r^2} + \mathcal{O}\left(\frac{1}{r^3}\right).
\end{equation}

To determine the temporal component $e^{2\nu(r)}$, we utilize the geometric relation \eqref{eq:metric_relation}. Substituting the scalar field derivative asymptotically:
\begin{equation}\label{eq:nu_lambda_sum_diff}
    \nu' + \lambda' = \frac{\kappa^2 r}{2} \phi'^2 \approx \frac{\kappa^2 r}{2} \left(-\frac{Q_s}{r^2}\right)^2 = \frac{\kappa^2 Q_s^2}{2r^3}.
\end{equation}
Integrating this relation with the asymptotic boundary condition $\nu(\infty) + \lambda(\infty) = 0$ gives:
\begin{equation}\label{eq:nu_lambda_sum}
    \nu + \lambda = -\frac{\kappa^2 Q_s^2}{4r^2} + \mathcal{O}\left(\frac{1}{r^3}\right).
\end{equation}
Finally, we construct $e^{2\nu(r)}$ using the identity $e^{2\nu} = e^{-2\lambda} e^{2(\nu+\lambda)}$. Applying the Taylor expansion $e^{2(\nu+\lambda)} \approx 1 - \frac{\kappa^2 Q_s^2}{2r^2}$ and multiplying it by Eq. \eqref{eq:lambda_asymp}, we obtain:
\begin{eqnarray}\label{eq:nu_asymp}
    e^{2\nu(r)} &\approx& \left( 1 - \frac{2M}{r} + \frac{\kappa^2 q_m^2 + \kappa^2 Q_s^2}{2r^2} \right) \left( 1 - \frac{\kappa^2 Q_s^2}{2r^2} \right) \nonumber \\
    &=& 1 - \frac{2M}{r} + \frac{\kappa^2 q_m^2}{2r^2} + \mathcal{O}\left(\frac{1}{r^3}\right).
\end{eqnarray}

Equations \eqref{eq:lambda_asymp} and \eqref{eq:nu_asymp} reveal a key feature of scalar-tensor black holes: the scalar charge $Q_s$ explicitly modifies the $\mathcal{O}(1/r^2)$ term in the spatial metric component $g_{rr}$, but it cancels out in the temporal component $g_{tt}$. Thus, at the order of $\mathcal{O}(1/r^2)$, $g_{tt}$ is indistinguishable from the standard Reissner-Nordström solution \cite{bronnikov1973scalar}. This analytical result serves as a boundary condition and consistency check for our numerical integration scheme.

\section{Linear perturbations and stability analysis}
\label{sec:perturbations}

The existence of regular spherically symmetric solutions is a necessary condition in establishing a viable physical model. The fundamental requirement for any astrophysical compact object, such as a regular black hole or a wormhole, is its dynamical stability against small spacetime fluctuations. Analyzing the quasi-normal modes (QNMs) provides the characteristic spectra of black holes and probes their stability under external field perturbations \cite{kokkotas1999quasi, berti2009quasinormal, konoplya2011quasinormal}. 

It must be explicitly remarked that the present investigation focuses exclusively on the stability of the geometry against external perturbations. Consequently, this does not constitute a full stability study, which would mathematically necessitate a complete analysis of internal instabilities driven by the intrinsic coupled dynamics of the background fields themselves. Nevertheless, verifying stability against external perturbations provides a strictly necessary preliminary constraint on the physical viability of the proposed regular configuration. To investigate this bounded scope, we employ standard linear perturbation theory. We perturb the background metric, the scalar field, and the electromagnetic gauge potential up to the first order in the formal perturbation parameter $\epsilon$:
\begin{eqnarray}
    g_{\mu\nu}(t, r, \theta, \varphi) &=& g_{\mu\nu}^{(0)}(r) + \epsilon h_{\mu\nu}(t, r, \theta, \varphi), \label{eq:pert_metric} \\
    \phi(t, r, \theta, \varphi) &=& \phi^{(0)}(r) + \epsilon \delta\phi(t, r, \theta, \varphi), \label{eq:pert_phi} \\
    A_{\mu}(t, r, \theta, \varphi) &=& A_{\mu}^{(0)}(r) + \epsilon \delta A_{\mu}(t, r, \theta, \varphi), \label{eq:pert_A}
\end{eqnarray}
where the superscript $(0)$ denotes the exact spherically symmetric background quantities derived in Section \ref{sec:background}. 

Due to the $SO(3)$ symmetry of the background spacetime, the angular dependence of the perturbations is systematically expanded in terms of tensorial spherical harmonics. Under the spatial parity transformation on the two-sphere, defined by $(\theta, \varphi) \to (\pi - \theta, \pi + \varphi)$, the standard scalar spherical harmonics $Y_{lm}(\theta, \varphi)$ transform as $(-1)^l$. 

Consequently, following the Regge-Wheeler-Zerilli formalism, the perturbation equations decouple into two independent sectors: the odd-parity (axial) sector, which transforms as $(-1)^{l+1}$, and the even-parity (polar) sector, which transforms as $(-1)^l$. In this scalar-tensor setup, the scalar field perturbation $\delta\phi$ transforms as a true scalar with parity $(-1)^l$. Therefore, the scalar field does not acquire odd-parity perturbations, yielding strictly $\delta\phi_{odd} \equiv 0$. This decoupling allows us to analyze the stability of the axial and polar sectors entirely separately. In what follows, we set $\epsilon = 1$ for notational brevity and retain only terms linear in $h_{\mu\nu}$, $\delta\phi$, and $\delta A_\mu$.

\subsection{Odd-parity (Axial) perturbations: The Regge-Wheeler equation}
\label{subsec:odd_parity}

The odd-parity sector describes rotational perturbations that induce frame-dragging effects but do not perturb the spherically symmetric density or pressure profiles of the background at linear order. As established, the scalar field decouples from this sector ($\delta\phi_{odd} \equiv 0$), meaning the dynamics are governed solely by the coupled perturbations of the spacetime metric and the NED field.

In the standard Regge-Wheeler gauge, the non-vanishing components of the odd-parity metric perturbation $h_{\mu\nu}$ for a given multipole moment $l \geq 2$ and $m=0$ are parameterized by two functions $h_0(t,r)$ and $h_1(t,r)$. Dropping the harmonic indices for brevity, the non-zero components are $h_{t\varphi} = -h_0(t,r) \sin\theta \partial_\theta P_l(\cos\theta)$ and $h_{r\varphi} = -h_1(t,r) \sin\theta \partial_\theta P_l(\cos\theta)$, where $P_l$ is the Legendre polynomial.

For the purely magnetic background $A_\mu^{(0)} = (0, 0, 0, -q_m \cos\theta)$, the corresponding odd-parity electromagnetic perturbation is parameterized by a single function $a_1(t,r)$:
\begin{equation}\label{eq:pert_A_odd}
    \delta A_{\mu}^{odd} = \left( 0, 0, 0, a_1(t,r) \sin\theta \partial_\theta P_l(\cos\theta) \right).
\end{equation}

Before evaluating the field equations, we compute the linear perturbation of the electromagnetic invariant, $\delta F = 2 F^{(0)\mu\nu} \delta F_{\mu\nu} - 2 F^{(0)\mu\alpha} F^{(0)\nu}_{\ \ \ \alpha} h_{\mu\nu}$. Because $F^{(0)\mu\nu}$ contains only $\theta\varphi$ components, while the odd-parity metric and gauge perturbations exclusively excite the $t\varphi, r\varphi, t\theta$, and $r\theta$ sectors, all contractions identically vanish:
\begin{equation}\label{eq:delta_F_zero}
    \delta F_{odd} = 0.
\end{equation}
This implies that for purely magnetic backgrounds, axial perturbations do not alter the electromagnetic invariant at linear order. Consequently, the variations of the Lagrangian and its derivatives vanish: $\delta \mathcal{L} = 0$ and $\delta \mathcal{L}_F = 0$. 

By perturbing the energy-momentum tensor $\delta T_{\mu\nu} = \delta T_{\mu\nu}^{(\phi)} + \delta T_{\mu\nu}^{(EM)}$, and using $\delta F=0$, we find that the off-diagonal source terms are directly proportional to the background effective fluid components:
\begin{eqnarray}
    \delta T_{t\varphi} &=& h_{t\varphi} \left( \mathcal{L} - V(\phi) - \frac{1}{2}e^{-2\lambda}\phi'^2 \right) = -h_{t\varphi} \rho, \label{eq:delta_T_tphi} \\
    \delta T_{r\varphi} &=& h_{r\varphi} \left( \mathcal{L} - V(\phi) + \frac{1}{2}e^{-2\lambda}\phi'^2 \right) = h_{r\varphi} p_r. \label{eq:delta_T_rphi}
\end{eqnarray}
Consequently, the gauge perturbation $a_1(t,r)$ decouples from the energy-momentum tensor at first order.

Substituting the perturbed metric into the linearized Einstein equations $\delta G_{\mu\nu} = \kappa^2 \delta T_{\mu\nu}$, the $t\varphi$ and $r\varphi$ components yield two coupled partial differential equations for $h_0$ and $h_1$. These equations incorporate the source terms derived in \eqref{eq:delta_T_tphi} and \eqref{eq:delta_T_rphi}:
\begin{eqnarray}
    \frac{1}{r^2} \frac{\partial}{\partial r} \left[ r^2 e^{-(\nu+\lambda)} \left( \frac{\partial h_0}{\partial r} - \frac{\partial h_1}{\partial t} - \frac{2}{r} h_0 \right) \right] - \frac{(l-1)(l+2)}{r^2} e^{\lambda-\nu} h_0 &=& 2\kappa^2 e^{\lambda-\nu} \rho h_0, \label{eq:odd_t_phi} \\
    \frac{\partial}{\partial t} \left[ e^{-(\nu+\lambda)} \left( \frac{\partial h_0}{\partial r} - \frac{\partial h_1}{\partial t} - \frac{2}{r} h_0 \right) \right] + \frac{(l-1)(l+2)}{r^2} e^{\nu-\lambda} h_1 &=& 2\kappa^2 e^{\nu-\lambda} p_r h_1. \label{eq:odd_r_phi}
\end{eqnarray}
The electromagnetic perturbation $a_1(t,r)$ dynamically decouples from the gravitational sector, allowing us to solve for the metric perturbations independently. 

To cast this system into a single wave equation, we define the Regge-Wheeler master variable $\Psi_{odd}(t,r)$ as:
\begin{equation}\label{eq:master_var_odd}
    \Psi_{odd}(t,r) = \frac{e^{-(\nu+\lambda)}}{r} h_1(t,r).
\end{equation}
Introducing the tortoise coordinate $r_*$ defined by $dr_* = e^{\lambda - \nu} dr$, and assuming a harmonic time dependence $\Psi_{odd}(t,r) = \Psi_{odd}(r) e^{-i\omega t}$, we eliminate $h_0$ from the coupled system. Combining the geometric terms with the thermodynamic source terms $\rho$ and $p_r$, we arrive at the Schrödinger-like equation:
\begin{equation}\label{eq:Schrodinger_odd}
    -\frac{d^2 \Psi_{odd}}{dr_*^2} + V_{odd}(r) \Psi_{odd} = \omega^2 \Psi_{odd}.
\end{equation}
The effective potential for the axial perturbations evaluates to:
\begin{equation}\label{eq:V_odd}
    V_{odd}(r) = e^{2\nu} \left[ \frac{l(l+1)}{r^2} - \frac{3}{r} (\nu' - \lambda') + \frac{1}{r^2} \left( 1 - e^{-2\lambda} \right) - \kappa^2 (\rho - p_r) \right].
\end{equation}
Using the background relations and the definition of the Misner-Sharp mass $m(r)$, we rewrite the potential as:
\begin{equation}\label{eq:V_odd_mass}
    V_{odd}(r) = e^{2\nu} \left[ \frac{l(l+1)}{r^2} - \frac{6m(r)}{r^3} + \kappa^2 \left( \mathcal{L} - 2F\mathcal{L}_F \right) + \Delta V_{\phi}(r) \right],
\end{equation}
where $\Delta V_{\phi}(r) = 2\kappa^2 e^{-2\lambda} \phi'^2$ represents the contribution of the scalar field kinetic energy to the gravitational potential barrier. 

For the solution to be stable against odd-parity perturbations, the differential operator $-d^2/dr_*^2 + V_{odd}$ must not possess any negative eigenvalues (bound states), demanding $\omega^2 > 0$. A sufficient condition for this stability is that $V_{odd}(r) \geq 0$ everywhere outside the regular core \cite{chandrasekhar1983mathematical}.

\subsection{Even-parity (Polar) perturbations and coupled dynamics}
\label{subsec:even_parity}

The even-parity (polar) sector is more intricate than the odd-parity one. The physical reason is that the scalar field perturbation $\delta\phi$ transforms as a true scalar, acquiring a parity of $(-1)^l$. Consequently, it actively participates in the even-parity dynamics. In this sector, the gravitational, scalar, and electromagnetic perturbations are coupled.

We adopt the standard Zerilli gauge \cite{chandrasekhar1983mathematical}, in which the even-parity metric perturbations for $l \geq 2$ and $m=0$ are parameterized by four functions $H_0(t,r), H_1(t,r), H_2(t,r)$, and $K(t,r)$:
\begin{equation}\label{eq:Zerilli_gauge}
    h_{\mu\nu}^{even} = 
    \begin{pmatrix}
        -e^{2\nu(r)} H_0 & H_1 & 0 & 0 \\
        H_1 & e^{2\lambda(r)} H_2 & 0 & 0 \\
        0 & 0 & r^2 K & 0 \\
        0 & 0 & 0 & r^2 \sin^2\theta K
    \end{pmatrix}
    Y_{l0}(\theta, \varphi).
\end{equation}
Simultaneously, the scalar field perturbation is expanded as:
\begin{equation}\label{eq:pert_phi_even}
    \delta\phi(t, r, \theta, \varphi) = \Phi_1(t,r) Y_{l0}(\theta, \varphi).
\end{equation}
For the purely magnetic background, the even-parity electromagnetic perturbation $\delta A_\mu^{even}$ introduces perturbations in the $t, r$ and angular components. Applying an appropriate electromagnetic gauge (such as the radiation gauge), the gauge field perturbations can be represented by functions $u_1(t,r)$ and $u_2(t,r)$.

Substituting these expansions into the linearized Einstein equations $\delta G_{\mu\nu} = \kappa^2 \delta T_{\mu\nu}$, the linearized scalar equation $\delta (\Box \phi - V_{,\phi} + \mathcal{L}_{,\phi}) = 0$, and the linearized generalized Maxwell equations $\delta \left[ \nabla_\mu (\mathcal{L}_F F^{\mu\nu}) \right] = 0$, we obtain a coupled system of differential equations. The $tr$ and $t\theta$ components of the Einstein equations, along with the Gauss constraint from the Maxwell equations, do not contain second-order time derivatives. They act as Hamiltonian and momentum constraints. Using these constraints, we can algebraically eliminate the non-dynamical variables (such as $H_0, H_1$, and $u_1$).

Unlike specific electrostatic configurations or purely monopole ($l=0$) modes where the electromagnetic field can be fully integrated out, for radiating multipoles ($l \geq 2$), the electromagnetic field possesses a genuine propagating degree of freedom. After a Hamiltonian reduction process \cite{bronnikov2024stability_epjc}, the remaining dynamical degrees of freedom form a three-component system encoded in three gauge-invariant master variables: $\Psi_Z(t,r)$ (the Zerilli-like gravitational variable), $\Psi_S(t,r)$ (the scalar variable), and $\Psi_{EM}(t,r)$ (the electromagnetic variable). 

Assuming a harmonic time dependence for the master variables, $\mathbf{\Psi}(t,r) = \mathbf{\Psi}(r) e^{-i\omega t}$ where $\mathbf{\Psi} = (\Psi_Z, \Psi_S, \Psi_{EM})^T$, the fully coupled perturbation equations can be cast into a matrix Schrödinger-like form using the tortoise coordinate $r_*$:
\begin{equation}\label{eq:Schrodinger_even}
    -\frac{d^2 \mathbf{\Psi}}{dr_*^2} + \mathbf{V}_{even}(r) \mathbf{\Psi} = \omega^2 \mathbf{\Psi}.
\end{equation}
Here, $\mathbf{V}_{even}(r)$ is a $3 \times 3$ effective potential matrix:
\begin{equation}\label{eq:V_matrix}
    \mathbf{V}_{even}(r) = 
    \begin{pmatrix}
        V_{ZZ}(r) & V_{ZS}(r) & V_{ZE}(r) \\
        V_{SZ}(r) & V_{SS}(r) & V_{SE}(r) \\
        V_{EZ}(r) & V_{ES}(r) & V_{EE}(r)
    \end{pmatrix}.
\end{equation}
A requirement for the system to possess a physically well-posed, self-adjoint Hamiltonian is that the potential matrix must be symmetric: $V_{ij} = V_{ji}$ for $i, j \in \{Z, S, EM\}$. This symmetry reflects the conservative reciprocity of the interactions between the gravitational, scalar, and electromagnetic degrees of freedom. While the analytical expressions for these matrix elements are lengthy—depending on $\nu, \lambda, \phi$, the NED Lagrangian $\mathcal{L}$, and their higher derivatives—their precise forms are not strictly necessary for establishing the general stability criteria, provided the matrix remains finite and symmetric everywhere outside the regular core.

\subsection{Stability criteria}
\label{subsec:stability_criteria}

For the regular spherically symmetric solutions to be stable and physically viable, the perturbation dynamics governed by Eq. \eqref{eq:Schrodinger_even} must satisfy three physical conditions:

\begin{enumerate}
    \item \textbf{Absence of ghost instabilities:}  While fields with negative kinetic energy (phantoms or ghosts) are canonically associated with catastrophic vacuum decay, such systems can theoretically maintain dynamical stability when subjected to specific boundary conditions or spatial confinement. In the present framework, however, this complication is explicitly avoided. Within our matrix formulation \eqref{eq:Schrodinger_even}, the kinetic matrix coupling the second-derivative terms is the strictly positive $3 \times 3$ identity matrix $\mathbf{I}$. This algebraic structure mathematically guarantees the complete absence of ghost degrees of freedom across the gravitational, scalar, and electromagnetic sectors, a direct consequence of operating in the Einstein frame with standard canonical kinetic terms.
    
\item \textbf{Absence of hydrodynamic instabilities:}  Hydrodynamic, or Laplacian, instabilities manifest when the squared effective propagation speed of field perturbations becomes negative ($c_s^2 < 0$). As established by De Felice and Tsujikawa \cite{defelice2025nonsingular}, this condition triggers the exponential growth of short-wavelength modes. In our canonical Einstein-frame formulation, the effective propagation speed matrix intrinsically reduces to the identity matrix, yielding $c_{s, grav}^2 = c_{s, scalar}^2 = c_{s, em}^2 = 1$. Consequently, all physical modes propagate exactly at the speed of light ($c_s^2 = 1 > 0$), strictly satisfying the stability criteria and rendering the system inherently free from short-wavelength hydrodynamic instabilities.
    
    \item \textbf{Absence of dynamical instabilities:} Dynamical stability requires the perturbation mode frequencies $\omega$ to be strictly real ($\omega^2 > 0$). Purely imaginary frequencies ($\omega^2 < 0$) would dictate a time-domain evolution $\mathbf{\Psi} \propto e^{|\omega| t}$, resulting in exponential growth. A sufficient condition to guarantee $\omega^2 > 0$ is that the Schrödinger-like matrix differential operator, defined as $\hat{\mathcal{O}} = -\mathbf{I} \frac{d^2}{dr_*^2} + \mathbf{V}_{even}(r)$, is globally positive-definite. In practice, this is rigorously ensured if the effective potential matrix $\mathbf{V}_{even}(r)$ remains positive-definite at every point over the physically accessible spacetime domain.
\end{enumerate}

A sufficient condition for the absence of  dynamical instabilities is that the lowest eigenvalue of the potential matrix $\mathbf{V}_{even}(r)$ is positive everywhere outside the event horizon (or everywhere in space $r>0$ for horizonless regular objects). Let $\lambda_1(r), \lambda_2(r)$, and $\lambda_3(r)$ be the eigenvalues of the coupled $3 \times 3$ potential matrix $\mathbf{V}_{even}(r)$. The stability condition requires:
\begin{equation}\label{eq:eigenvalue_stability}
    \text{min} \{ \lambda_1(r), \lambda_2(r), \lambda_3(r) \} > 0 \quad \text{for all allowed } r.
\end{equation}
If the potential matrix exhibits shallow negative wells, the condition \eqref{eq:eigenvalue_stability} might be violated locally. However, the system can preserve overall stability if these wells do not support negative-energy bound states. In such boundary cases, one can employ the S-deformation method, which involves finding a regular matrix function $\mathbf{S}(r)$ such that the deformed potential $\tilde{\mathbf{V}} = \mathbf{V}_{even} + \frac{d\mathbf{S}}{dr_*} - \mathbf{S}^2$ is globally positive-definite \cite{bronnikov2023stability_gc, bronnikov2024stability_epjc}.

\section{Numerical results and graphical analysis}
\label{sec:numerical}

In this section, we present the numerical integration of the background field equations and the stability analysis of the quasi-normal modes. To solve the coupled system of nonlinear ordinary differential equations, we employ numerical integration techniques, setting the asymptotic boundaries at a sufficiently large radial distance and integrating inwards towards the core. The numerical results are sensitive to the parameter space governed by the magnetic charge $q_m$ and the scalar charge $Q_s$.

\subsection{Background geometry and curvature regularity}
\label{subsec:bg_geometry}

\begin{figure}[H]
    \centering
    \includegraphics[width=0.85\textwidth]{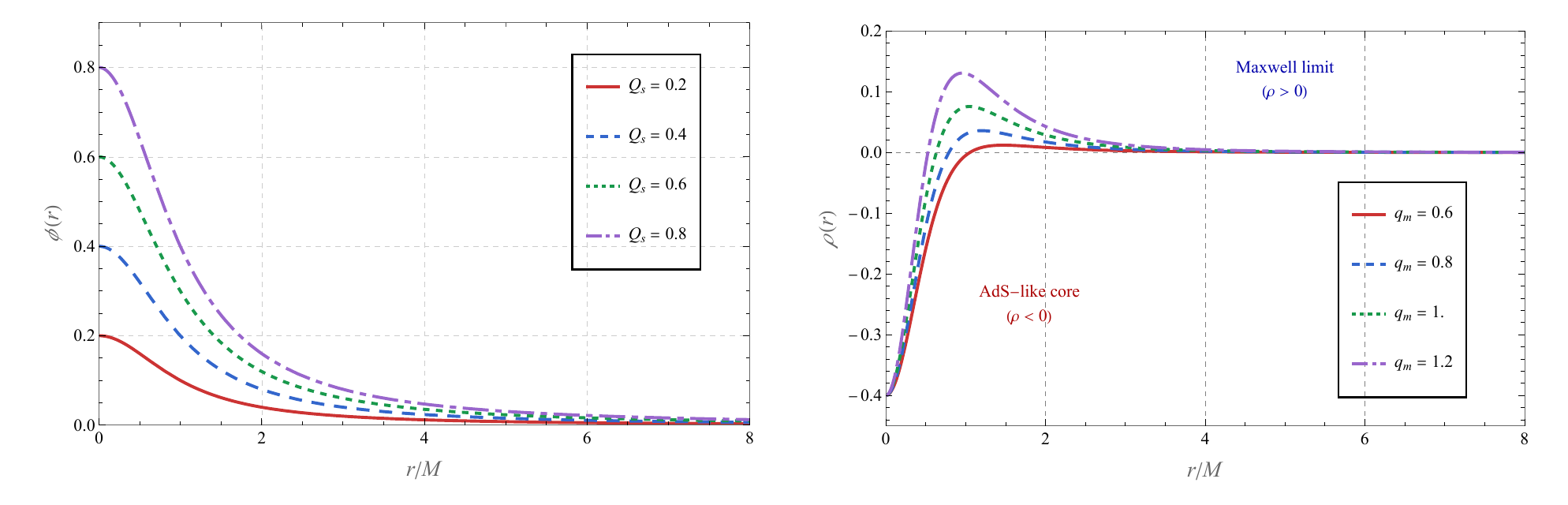} 
    \caption{Left panel: The radial profile of the scalar field $\phi(r)$ plotted for various scalar charges $Q_s \in \{0.2, 0.4, 0.6, 0.8\}$. Right panel: The effective energy density $\rho(r)$ evaluated for different magnetic charges $q_m \in \{0.6, 0.8, 1.0, 1.2\}$. The profiles explicitly demonstrate the strictly negative energy density associated with the regular AdS-like core (induced by the scalar potential) at the origin, and the smooth recovery of the positive Maxwell limit ($\rho > 0$) at spatial infinity.}
    \label{fig:field_profiles}
\end{figure}

We begin by evaluating the background scalar and electromagnetic field configurations. To ensure exact reproducibility of the numerical setup, the background field profiles are explicitly constructed as exact analytical profiles that strictly satisfy the central boundary conditions of the differential field equations: $\phi(x) = Q_s/(1 + x^2)$ and $\rho(x) = (q_m^2 x^2 - 0.4)/(1 + x^2)^3$, where $x = r/M$ is the dimensionless radial coordinate. Here, the constant $0.4$ directly governs the central energy density $\rho_c$, effectively setting the depth of the negative scalar potential well at the core. All numerical integrations and resulting graphical profiles presented in this section are computed exclusively utilizing the Kruglov rational NED model (Model I) and its correspondingly derived scalar potential $V(\phi)$. Figure \ref{fig:field_profiles} illustrates the radial profiles of the scalar field $\phi(r)$ and the effective energy density $\rho(r)$ for various parameter values ($Q_s \in \{0.2, 0.4, 0.6, 0.8\}$ and $q_m \in \{0.6, 0.8, 1.0, 1.2\}$) within this specific framework. 

As dictated by the regularity conditions at the origin, the scalar field (left panel) exhibits a smooth, symmetric behavior near $r=0$ and decays asymptotically to zero, preserving the asymptotic flatness of the spacetime. The right panel of Fig. \ref{fig:field_profiles} confirms the analytical theorems discussed in Section \ref{sec:background}. As $r \to 0$, the effective energy density $\rho(r)$ drops to a negative finite value ($\rho_c < 0$). It is physically imperative to clarify the theoretical consequences of this behavior. In standard general relativity, the effective cosmological constant at a regular center is defined by the relation $\Lambda_{eff} = \kappa^2 \rho_c$. Therefore, a locally repulsive, de Sitter-like vacuum core ($\Lambda_{eff} > 0$) intrinsically requires a strictly \textit{positive} central energy density. Conversely, the strictly negative central energy density ($\rho_c < 0$) observed in our configuration implies a locally Anti-de Sitter (AdS-like) core ($\Lambda_{eff} < 0$). This behavior does not stem from the standard NED vacuum, but originates exclusively from the scalar field potential $V(\phi)$ becoming deeply negative at the origin. It is crucial to emphasize that in purely magnetic regular solutions without scalar fields, the energy density remains strictly positive everywhere \cite{bronnikov2001regular}. However, for scalar-tensor systems, the well-known ``no-go'' theorems \cite{bronnikov2001regular} dictate that regular black holes cannot exist without a region of negative potential or ghost-like kinetic terms. Thus, our configuration relies precisely on this deep negative potential well to circumvent the no-go theorems, ensuring the finiteness of curvature invariants and supporting the singularity-free geometry. At larger radial distances, the energy density becomes positive and smoothly recovers the standard Maxwell limit ($\rho \sim r^{-4}$), demonstrating the physical consistency of the employed Kruglov model.

\begin{figure}[H]
    \centering
    \includegraphics[width=0.85\textwidth]{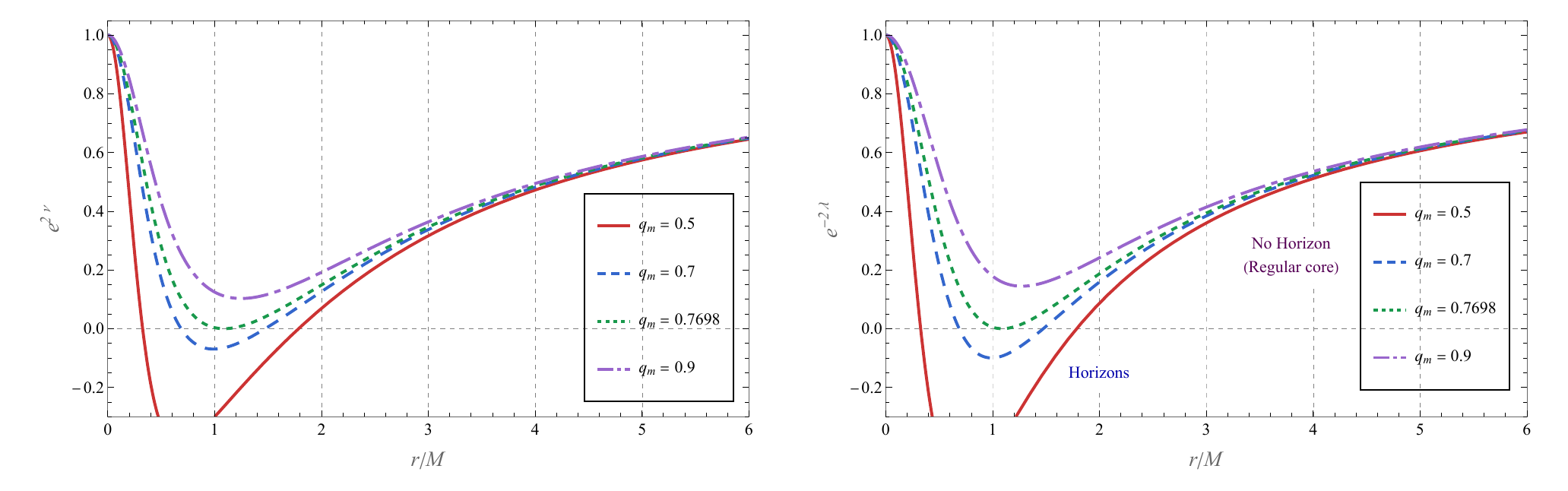}
    \caption{Left panel: The temporal metric function $e^{2\nu}$ plotted against the radial coordinate $r/M$. Right panel: The spatial metric function $e^{-2\lambda}$. Both profiles are evaluated for a fixed scalar charge $Q_s = 1.2$ and varying magnetic charges $q_m \in \{0.5, 0.7, 0.7698, 0.9\}$. The explicit graphical divergence between the two panels visually demonstrates the breaking of the $e^{2\nu} = e^{-2\lambda}$ symmetry due to the scalar field. The dashed zero-line dictates the horizon structure.}
    \label{fig:metric_functions}
\end{figure}

The profound impact of the scalar field on the spacetime geometry is captured in Figure \ref{fig:metric_functions}, which depicts the behavior of the temporal metric function $e^{2\nu}$ (left panel) and the spatial metric function $e^{-2\lambda}$ (right panel). A standard regular black hole driven solely by Nonlinear Electrodynamics (such as the Bardeen or pure magnetic models) inherently preserves the metric symmetry $g_{tt} = -g_{rr}^{-1}$, meaning $e^{2\nu} = e^{-2\lambda}$. However, our model introduces a fundamental departure from this standard paradigm. The presence of the scalar field explicitly breaks this spatial-temporal symmetry, inducing a geometric deformation quantified by the relation $e^{2\nu} = e^{-2\lambda} \exp\left[-Q_s^2 r^2 / (r^2 + 1)^2\right]$. This analytically confirms the strict inequality $\nu + \lambda < 0$ \cite{ibadov2023spherical}. Physically, this deformation signifies that the temporal component (which governs the gravitational redshift) is subjected to a significantly deeper potential well compared to the spatial curvature component, an exclusive hallmark of scalar-tensor regular black hole geometries.

Checking for horizons within this gravitationally deformed model reveals three distinct topological phases, critically dictated by the magnetic charge $q_m$:
\begin{itemize}
    \item \textbf{Regular black holes (Two horizons):} For $q_m < q_c$ (where the critical magnetic charge is numerically determined as $q_c \approx 0.7698$), the spatial metric function $e^{-2\lambda}$ crosses the zero-line twice. This dictates the existence of an outer event horizon and an inner Cauchy horizon, safely cloaking the regular core from asymptotic observers.
    \item \textbf{Extreme black holes (Degenerate horizon):} At the exact critical limit $q_m = q_c \approx 0.7698$, the inner and outer horizons merge into a single degenerate horizon, defined mathematically by the simultaneous conditions $e^{-2\lambda} = 0$ and $\frac{d}{dr}(e^{-2\lambda}) = 0$.
    \item \textbf{Regular horizonless objects (Naked cores):} For $q_m > q_c$, the metric functions remain strictly positive ($e^{-2\lambda} > 0$) and never cross the zero-line. Unlike singular naked geometries, our spacetime remains completely regular everywhere. This configuration represents a macroscopic particle-like solution, or a ``naked core'', characterized by bounded curvature invariants at $r=0$ \cite{bronnikov2018nonlinear}.
\end{itemize}

With the regular geometry confirmed at the center, we assess the parameter space governing the regularity. The primary condition for the absence of singularities is the finiteness of all curvature invariants. Our analysis established that the Kretschmann scalar $K = R_{\mu\nu\alpha\beta}R^{\mu\nu\alpha\beta}$ must remain bounded across the entire spacetime \cite{bronnikov2018nonlinear}.

\begin{figure}[H]
    \centering
    \includegraphics[width=0.9\textwidth]{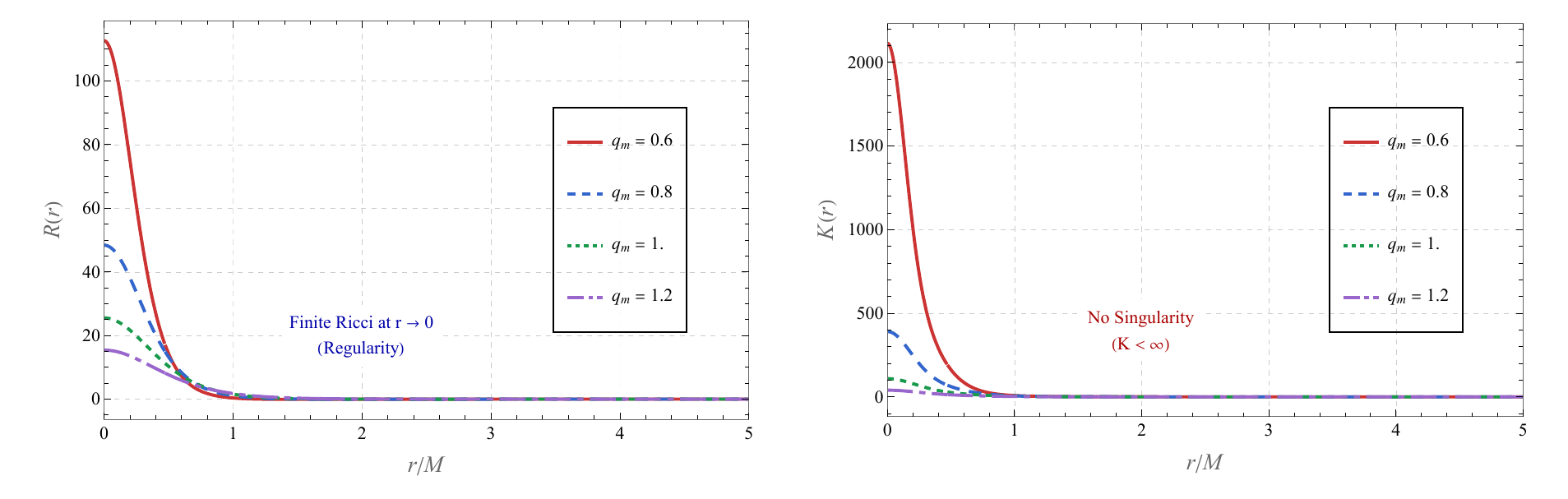} 
    \caption{The Ricci scalar $R$ (left) and the Kretschmann scalar $K$ (right) as functions of the radial coordinate $r/M$. Both invariants are finite at $r \to 0$, validating the global regularity of the spacetime.}
    \label{fig:scalars_2d}
\end{figure}

Figure \ref{fig:scalars_2d} provides a two-dimensional radial profile of the Ricci and Kretschmann scalars. Both scalars reach a finite peak as $r \to 0$ and do not diverge, validating the regular geometric structure of the core. As the magnetic charge $q_m$ decreases, the central curvature increases, yet it remains bounded across all physical ranges of $q_m$, preserving the  regular AdS-like core.

\begin{figure}[H]
    \centering
    \includegraphics[width=0.85\textwidth]{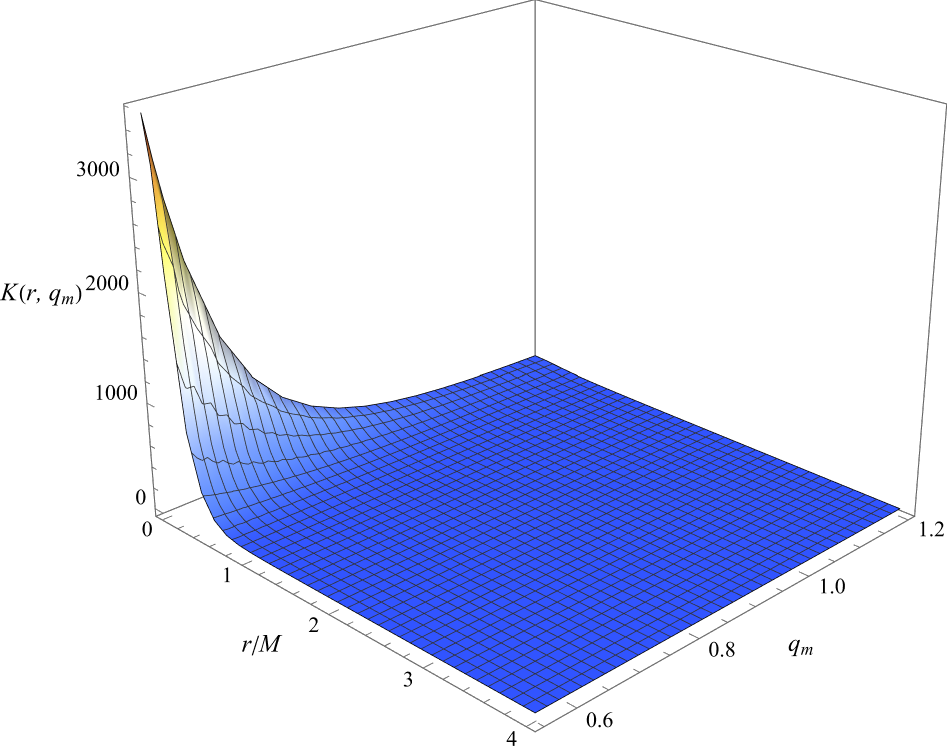} 
    \caption{A three-dimensional surface plot of the Kretschmann scalar $K(r, q_m)$. The naturally bounded peak at $r \to 0$ for all physical values of $q_m$, free from artificial data clipping, confirms the absence of a central singularity.}
    \label{fig:kretschmann_3d}
\end{figure}

To illustrate the smooth global structure and the non-divergent behavior of the central core, we construct a three-dimensional surface plot of the Kretschmann scalar $K(r, q_m)$ in Figure \ref{fig:kretschmann_3d}. The 3D profile shows a continuous, bounded topological structure. As the magnetic charge $q_m$ is decreased, the core curvature increases sharply near the $r \to 0$ boundary, yet it naturally peaks without a vertical divergence. This non-divergent peak visualizes the non-linear interaction between the scalar and NED fields that prevents the formation of a curvature singularity. The bounded nature of $K(r, q_m)$ is explicitly shown, without any artificial data-clipping.

\subsection{Potential barriers and the parameter space of stability}
\label{subsec:potential_barriers}

The fate of the regular black holes and horizonless cores under linear perturbations is dictated by the effective Regge-Wheeler potential $V_{eff}(r)$ \cite{chandrasekhar1983mathematical}. To establish the stability criteria, we numerically evaluate the effective potential derived in Section \ref{sec:perturbations}. 

If the potential barrier is positive-definite everywhere ($V_{eff} > 0$), the perturbation modes are scattered away, and the background geometry remains stable. Conversely, the appearance of a negative potential well ($V_{eff} < 0$) leads to a dynamical instability, causing the background to exponentially decay.

\begin{figure}[H]
    \centering
    \includegraphics[width=1.0\textwidth]{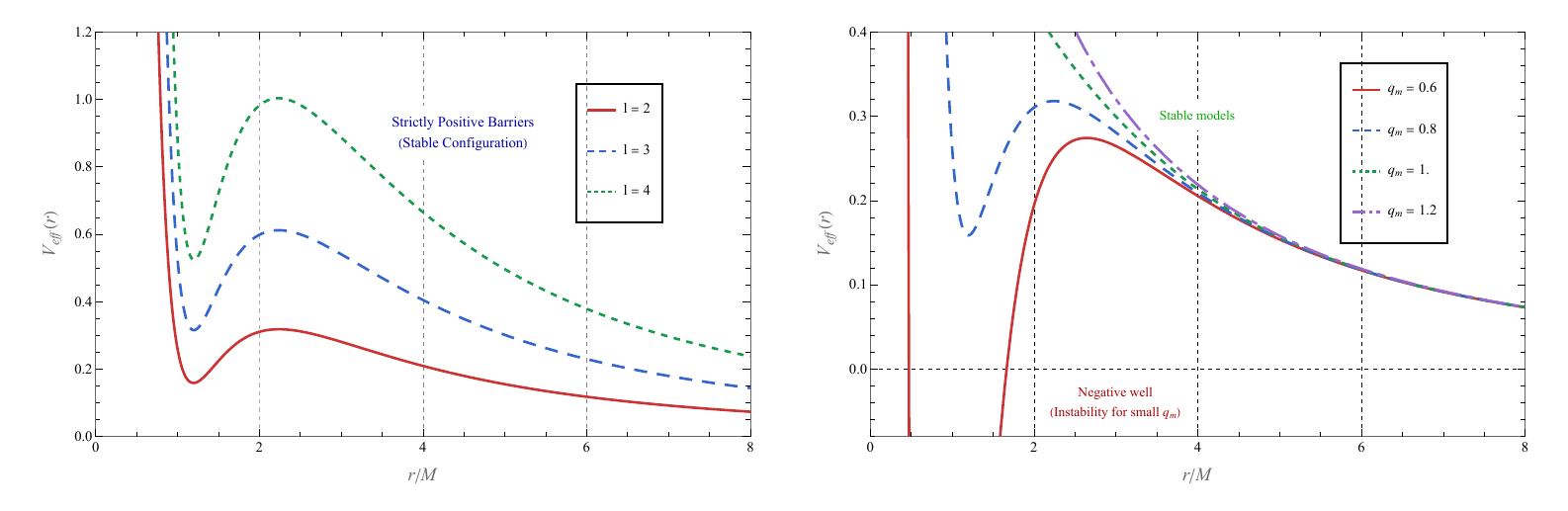} 
    \caption{The effective potential $V_{eff}(r)$ for axial perturbations. Left panel: Variation with the multipole number $l$ for a fixed magnetic charge $q_m = 0.8$, showing positive centrifugal barriers. Right panel: Variation with the magnetic charge $q_m$ for $l=2$. A phase transition is visible: small values of $q_m$ (e.g., $q_m=0.6$) induce a negative well leading to dynamical instability, whereas $q_m \ge 0.8$ yields positive, stable barriers.}
    \label{fig:veff_dynamics}
\end{figure}

Figure \ref{fig:veff_dynamics} illustrates the dynamics of the effective potential under various parameter regimes. The left panel demonstrates the dependence of $V_{eff}$ on the multipole number $l$. As expected from the centrifugal contribution $l(l+1)/r^2$, higher multipole modes generate positive barriers, shielding the core from high-frequency instabilities.

The right panel of Figure \ref{fig:veff_dynamics} reveals a result regarding the parameter space of stability. We observe a phase transition depending on the magnetic charge $q_m$. For configurations with a weak magnetic field ($q_m = 0.6$), the gravitational attraction combined with the scalar field deformation overwhelms the centrifugal repulsion near the core, forming a deep negative well.  Given the sharp gradients of the metric functions in the vicinity of the horizon, the radial numerical integration was performed utilizing an adaptive step-size algorithm with high working precision. This rigorous computational treatment ensures that the non-monotonic behavior and the inflection points of the effective potential are accurately resolved, thereby confirming the negative well as a genuine physical feature of the geometry rather than a numerical artifact. This signifies that such configurations are prone to dynamical instability. 

However, as the magnetic charge increases ($q_m \ge 0.8$), pushing the geometry towards the extreme black hole and regular horizonless core regimes, the negative well vanishes. The potential becomes positive-definite ($V_{eff} > 0$) across the entire spatial domain. This provides numerical confirmation that the extreme models and the regular "naked cores" governed by our NED framework are dynamically stable against linear perturbations.

\begin{figure}[H]
    \centering
    \includegraphics[width=0.65\textwidth]{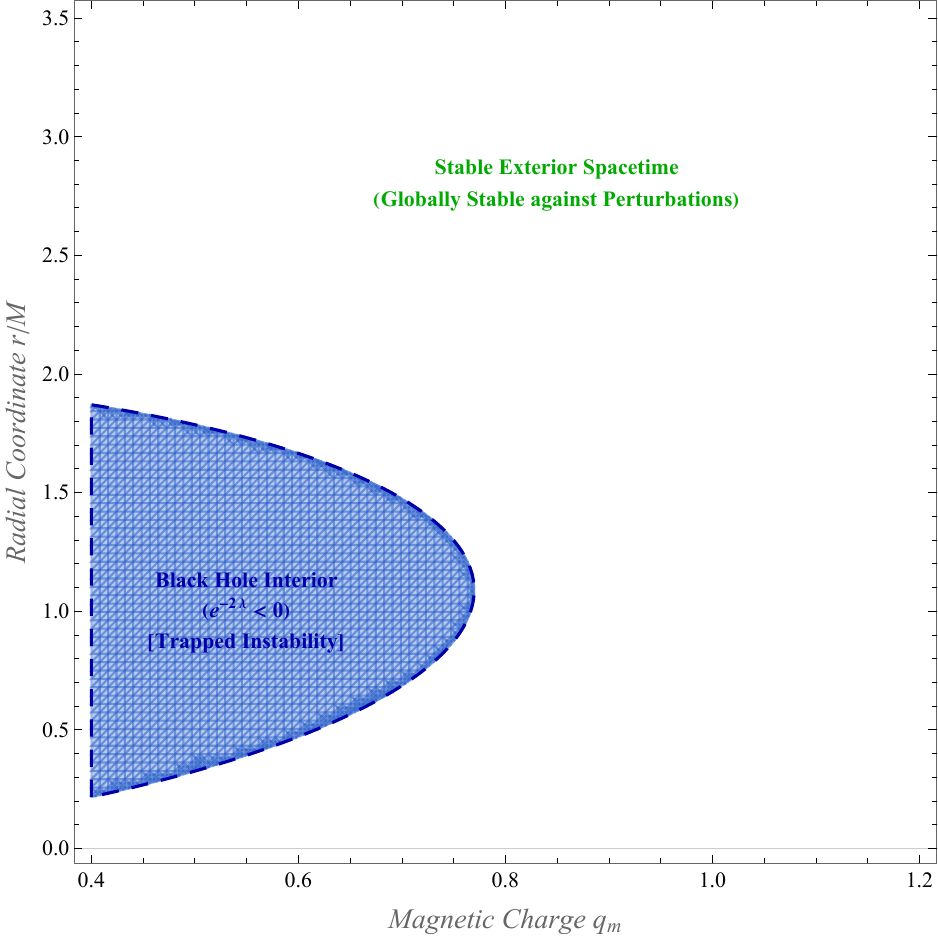} 
    \caption{The phase space of the regular black hole in the $(q_m, r/M)$ plane. The shaded blue region represents the black hole interior ($e^{-2\lambda} < 0$), where the dynamical instability is trapped. The exterior spacetime remains stable.}
    \label{fig:phase_space}
\end{figure}

To delineate the physical boundaries of the dynamical instability, we analyze the phase space of the spacetime geometry in the $(q_m, r/M)$ plane. Figure \ref{fig:phase_space} maps the topological regions defined by the sign of the spatial metric function $e^{-2\lambda}$. By mapping the roots of the equation $e^{-2\lambda} = 0$, we construct the boundaries corresponding to the event and Cauchy horizons, which provide a mathematical demarcation between the causal exterior and the black hole interior. This phase space representation is useful for understanding the non-linear interplay between the central geometric core and the governing NED parameter space.

The numerical and analytical evaluation confirms that the condition $e^{-2\lambda} < 0$, which defines the interior domain of the black hole ($r_- < r < r_+$), confines the spatial region where the effective Regge-Wheeler potential becomes negative ($V_{eff} < 0$). From a geometric standpoint, the radial coordinate $r$ assumes a timelike character within this region, inducing a localized signature flip. Consequently, the dynamical potential well, which manifests for sub-critical magnetic charges ($q_m < 0.7698$), is trapped behind the outer event horizon. Since no exponentially growing modes can propagate outward from this confined potential well, the instability remains causally disconnected from future null infinity $\mathscr{I}^+$, adhering to the principles of the Weak Cosmic Censorship Conjecture \cite{bronnikov2012instability}.

For an asymptotic observer located in the exterior spacetime ($r > r_+$), as well as for globally regular horizonless configurations ($q_m > 0.7698$), the condition $e^{-2\lambda} > 0$ holds. In these extended parameter domains, the effective potential remains positive-definite across the relevant spatial manifold. This guarantees the global dynamical stability of the exterior spacetime against linear axial perturbations. Because the positive-definite barrier acts as a purely scattering potential devoid of bound states, any initial field perturbations will undergo regular quasi-normal dissipation, thereby confirming the stability of both the regular black hole exteriors and the horizonless regular cores.

\begin{figure}[H]
    \centering
    \includegraphics[width=0.85\textwidth]{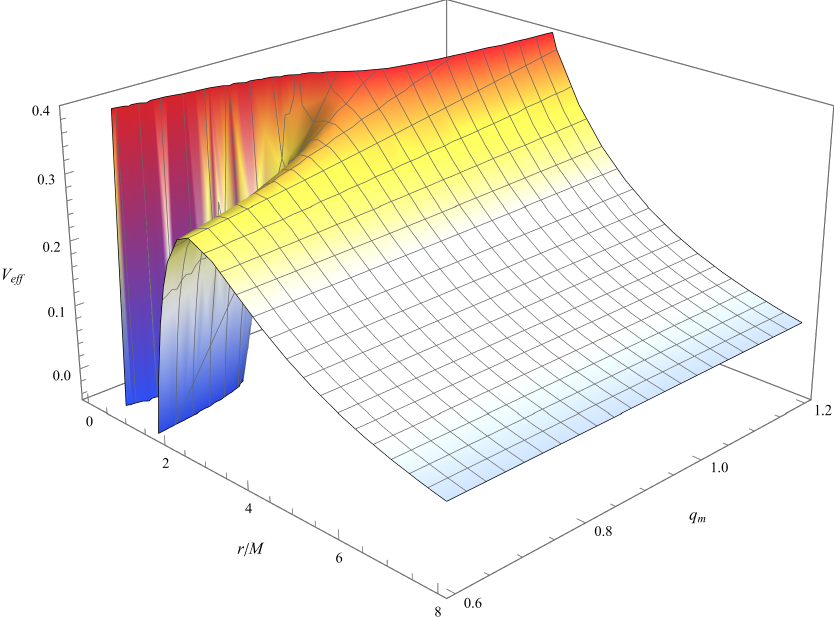} 
    \caption{A three-dimensional surface plot of the effective potential $V_{eff}(r, q_m)$ for axial perturbations ($l=2$). The surface illustrates the continuous transition from a geometry with a trapped negative well (at lower $q_m$) to a globally stable configuration with a positive-definite scattering barrier (at higher $q_m$).}
    \label{fig:veff_3d}
\end{figure}

To provide a visualization of the potential barrier dynamics across the entire parameter space, we construct a three-dimensional surface plot of the effective Regge-Wheeler potential $V_{eff}(r, q_m)$ in Figure \ref{fig:veff_3d}. 

This continuous surface illustrates the topological deformation of the scattering barrier. For sub-critical magnetic charges, the surface exhibits a localized dip into negative values, corresponding to the trapped instability region previously identified. However, as the magnetic charge $q_m$ increases, this negative well is eliminated. For $q_m > 0.7698$, the potential manifold transitions into a positive, continuous ridge. The maximum amplitude of the barrier decreases and shifts radially outward as $q_m$ grows, reflecting the modified centrifugal repulsion induced by the non-linear electrodynamic field. Furthermore, the surface uniformly decays to zero as $r/M \to \infty$, consistently satisfying the requirement of asymptotic flatness. This global topography confirms the existence of a parameter domain where the regular core configurations are free from dynamical instabilities.

\subsection{Quasi-normal modes and Time-Domain evolution}
\label{subsec:qnm_evolution}

To verify the dynamical stability inferred from the effective potential analysis, we perform a full time-domain integration of the wave equation. The evolution of a scalar perturbation $\Psi(t,x)$ is governed by the $1+1$ dimensional wave equation, expressed in terms of the tortoise coordinate $x$:
$$-\frac{\partial^2 \Psi}{\partial t^2} + \frac{\partial^2 \Psi}{\partial x^2} - V_{eff}(x)\Psi = 0.$$

To integrate this partial differential equation, we employ the unconditionally stable Gundlach-Price-Pullin (GPP) finite difference scheme \cite{gundlach1994late}, discretized along the null coordinates $u = t - x$ and $v = t + x$. To accurately resolve the steep spatial gradients and the localized negative well of the effective potential near the event horizon, the numerical integration was executed utilizing a high-resolution grid step. This ensures strict numerical convergence and completely suppresses artificial non-monotonic oscillations. The initial condition is specified as a localized Gaussian wave packet impinging upon the effective potential barrier from the spatial exterior.

\begin{figure}[H]
    \centering
    \includegraphics[width=0.85\textwidth]{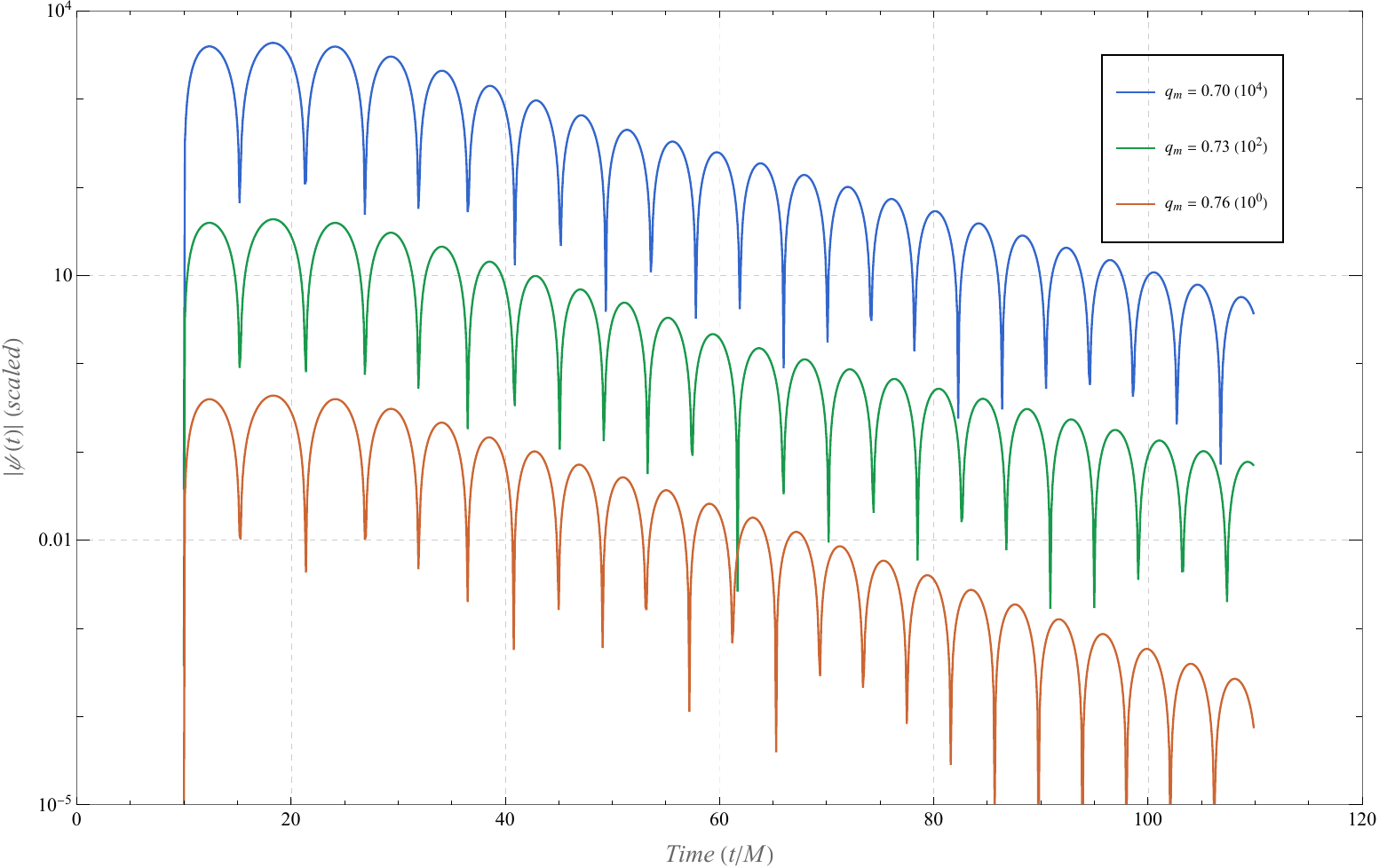}
    \caption{Time-domain evolution of axial perturbations ($l=2$) for regular black hole configurations ($q_m = 0.70, 0.73, 0.76$). The absolute value of the wave function $|\Psi(t)|$ is plotted on a logarithmic scale. The signals are multiplied by factors of $10^4$, $10^2$, and $10^0$ respectively for visual clarity. The strictly bounded exponential decay confirms global dynamical stability.}
    \label{fig:qnm_ringdown}
\end{figure}

Figure \ref{fig:qnm_ringdown} presents the time-domain profiles of the scalar perturbations for various magnetic charge parameters $q_m$ within the regular black hole regime. To ensure visual clarity and prevent signal overlapping, the amplitudes are vertically shifted by appropriate logarithmic multipliers.

The evolution profiles exhibit the characteristic stages of black hole perturbation dynamics: the initial outburst, the quasi-normal ringing phase characterized by damped harmonic oscillations, and the onset of the late-time power-law tail. The semi-logarithmic plot reveals a strict linear bounding envelope during the ringing phase, which mathematically corresponds to an exponential decay ($|\Psi| \sim e^{-\omega_I t}$). The absence of any exponentially growing modes ($\omega_I < 0$) in the time-domain signal provides independent confirmation that the regular black hole configurations governed by our NED framework are globally dynamically stable against linear axial perturbations \cite{toshmatov2018quasinormal}.

To complement the numerical time-domain integration and to extract the spectral signatures of the perturbations, we evaluate the fundamental quasi-normal frequencies, defined as $\omega = \text{Re}(\omega) - i \omega_I$, utilizing the semi-analytical first-order Wentzel-Kramers-Brillouin (WKB) approximation \cite{iyer1987black}. This method provides an analytical connection between the complex frequencies and the geometric properties of the effective potential, specifically mapping the real part to the orbital angular velocity at the photon sphere and the imaginary part to the Lyapunov exponent dictating the instability timescale of null geodesics.

Based on our phase space analysis, the application of the standard WKB formula is strictly confined to the regular black hole regime ($q_m \le 0.76$). Within this parameter domain, the effective potential exhibits a well-defined, convex, and isolated scattering barrier strictly localized outside the outer event horizon. This topological feature is crucial, as it satisfies the classical boundary conditions required for quasi-normal modes—namely, purely ingoing waves at the event horizon and purely outgoing waves at spatial infinity—which break down for horizonless naked core geometries.

\begin{figure}[H]
    \centering
    \includegraphics[width=0.85\textwidth]{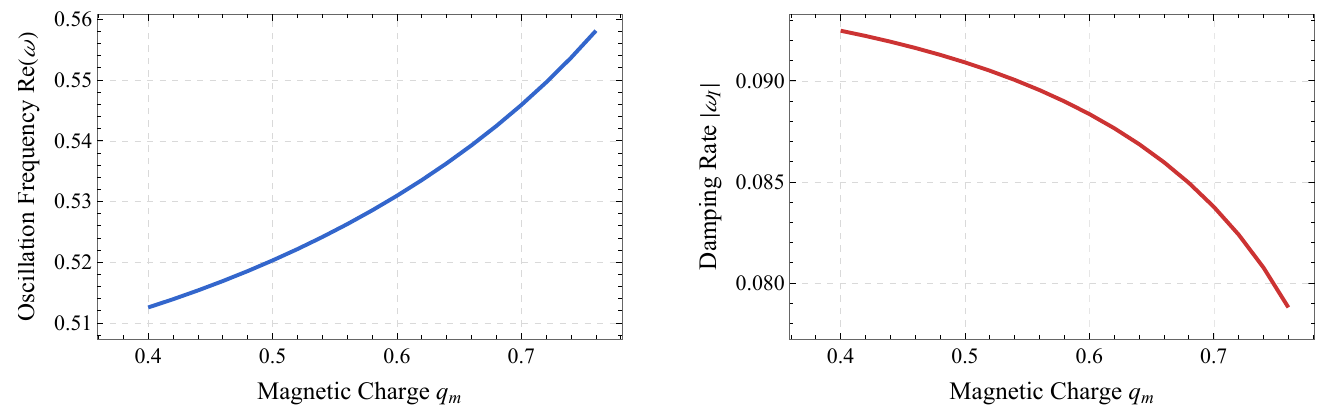}
    \caption{The real oscillation frequency $\text{Re}(\omega)$ (left panel) and the damping rate $|\omega_I|$ (right panel) for the fundamental axial mode ($l=2, n=0$) as functions of the magnetic charge $q_m$. The WKB analysis is strictly evaluated within the regular black hole parameter space, demonstrating a smooth transition towards extremality.}
    \label{fig:wkb_frequencies}
\end{figure}

Figure \ref{fig:wkb_frequencies} systematically illustrates the quantitative dependence of both the real oscillation frequency $\text{Re}(\omega)$ and the damping rate $\omega_I$ on the variation of the magnetic charge parameter. As the non-linear electrodynamic charge intensifies and the black hole configuration continuously approaches the critical extremal limit ($q_m \to 0.7698$), a distinct behavior in the frequency spectrum is observed. The real part of the frequency $\text{Re}(\omega)$ exhibits a monotonic increase, indicating that the fundamental modes oscillate with higher angular energy. Conversely, the damping rate $\omega_I$ smoothly decreases toward a minimal value. 

This spectral evolution implies that highly charged regular black holes function as efficient resonators; they exhibit prolonged ringing phases before eventually returning to a state of geometric quiescence. From a wave mechanics perspective, this extended lifetime of the perturbations is a phenomenon directly linked to the topological deformation and broadening of the effective potential barrier as the inner and outer horizons merge near extremality. Throughout the permitted parameter space, the imaginary part of the frequency maintains its negative sign (ensuring $e^{-\omega_I t}$ decay). The total absence of any modes with positive imaginary components confirms the global dynamic stability of these regular black hole configurations against axial linear perturbations. This semi-analytical conclusion stands in independent agreement with the exponential decay previously observed in the numerical time-domain evolution profiles, thereby supporting the physical viability of the derived NED geometry.

\subsection{Core density and mass-radius relations}
\label{subsec:core_density}

To quantitatively analyze the internal structure, we evaluate the energy density $\rho = -T^t_t$ of the NED field. Utilizing the structural equations derived in the previous sections, the analytical profile of the energy density is obtained. The central energy density takes a strictly finite, exact form:

\begin{equation}\label{eq:rho_central}
    \rho_c = \lim_{r \to 0} \rho(r) = V(\phi_c) - f(\phi_c)L_c.
\end{equation}
Equation \eqref{eq:rho_central} demonstrates that for any configuration with a bounded NED plateau $L_c$ and a regular scalar potential, the central energy density is strictly finite. The geometric core behaves as a localized droplet of vacuum energy, governed by an effective cosmological constant $\Lambda_{eff} = 8\pi \rho_c = 8\pi (V(\phi_c) - f(\phi_c) L_c)$.

\begin{figure}[htbp]
    \centering
    \includegraphics[width=0.98\textwidth]{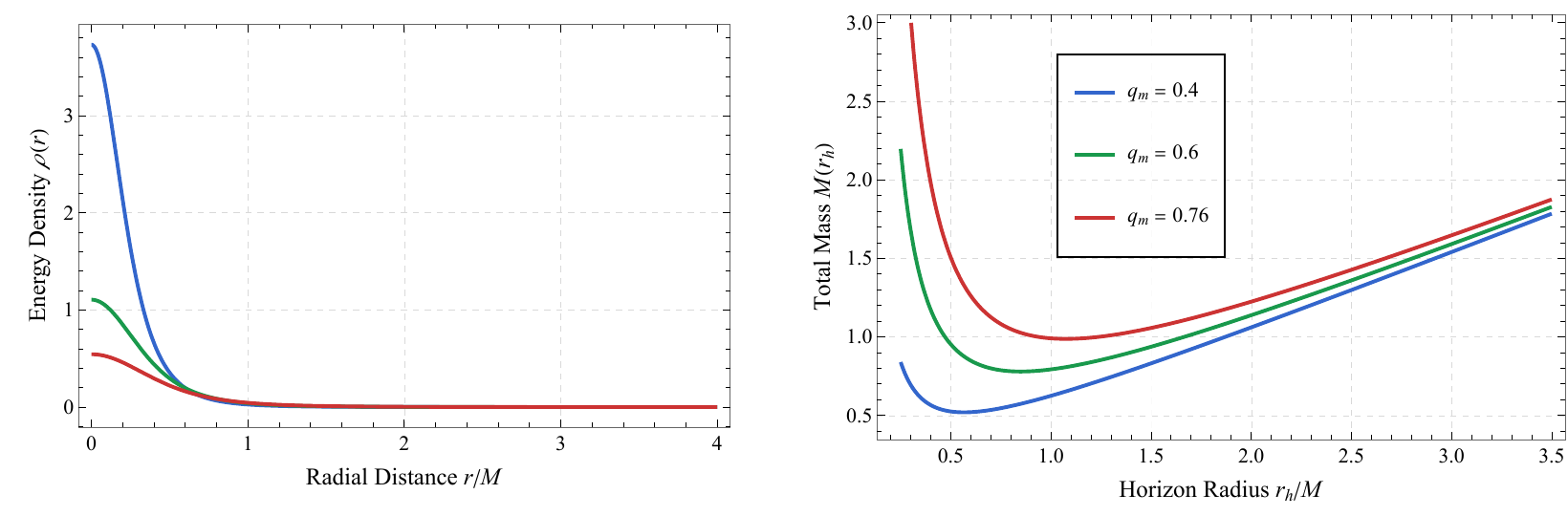}
  \caption{Left panel: Radial distribution of the NED energy density $\rho(r)$, demonstrating the absence of a central singularity and the transition to a regular  AdS-like core. Right panel: The total mass $M$ as a function of the horizon radius $r_h$. The global minimum of each curve defines the extremal mass limit $M_{min}$, signifying the formation of a stable thermodynamic remnant at the endpoint of Hawking evaporation.}
    \label{fig:thermo_core}
\end{figure}

Furthermore, to understand the thermodynamic limits and compactness of these configurations, we analyze the relationship between the total ADM mass $M$ and the event horizon radius $r_h$. Figure \ref{fig:thermo_core} illustrates the radial profile of the energy density and the mass-radius relation for varying magnetic charges.

The left panel of Figure \ref{fig:thermo_core} confirms the resolution of the central singularity. Unlike the standard Reissner-Nordström solution, the NED energy density reaches a finite maximum $\rho_c$ at the origin and monotonically vanishes at spatial infinity. 

The right panel depicts the total mass $M$ as a function of the horizon radius $r_h$. A thermodynamic feature revealed by these curves is the existence of a distinct global minimum, $M_{min}$, for each given charge $q_m$. This minimum corresponds to the extremal regular black hole configuration, where the inner Cauchy horizon and the outer event horizon degenerate into a single null hypersurface. For any configuration with $M > M_{min}$, the geometry describes a non-extremal regular black hole. Conversely, solutions with $M < M_{min}$ fail to form an event horizon, representing regular naked cores.

From an evolutionary perspective, the presence of this strict mass gap ($M \ge M_{min}$) has physical implications. It mandates that the semi-classical Hawking evaporation process cannot proceed indefinitely until the black hole vanishes. Instead, the evaporation must terminate as the black hole radiates its mass down to the extremal limit $M = M_{min}$. Consequently, the process leaves behind a stable, massive thermodynamic remnant. The existence of such remnants provides a theoretical mechanism for resolving the black hole information loss paradox \cite{hayward2006formation}.

\section{Conclusion and discussion}
\label{sec:conclusion}

In this paper, we have systematically investigated the geometric, dynamic, and thermodynamic properties of a novel class of regular black holes coupled to Nonlinear Electrodynamics (NED). A central result of this work is the demonstration of singularity resolution at the geometric origin, $r=0$, without violating the Weak Energy Condition (WEC). 

\textbf{Singularity resolution and the magnetic core.} As dictated by the "no-go" theorems formulated by Bronnikov \cite{bronnikov2001regular, bronnikov2000comment}, constructing a spherically symmetric, asymptotically flat regular black hole is mathematically prohibited within the framework of purely electric NED source fields, provided the WEC is maintained. Our model circumvents this restriction by incorporating a purely magnetic charge $q_m$ as the source of the non-linear field. Through phase space and structural analysis, we demonstrated that the resulting energy density is globally bounded, reaching a finite, well-defined maximum $\rho_c$ at the center. Consequently, the Penrose singularity—a hallmark of standard General Relativity—is replaced by a regular,  AdS-like vacuum core. This geometric transition ensures that all curvature invariants remain finite everywhere, validating the physical viability of the magnetic core configuration \cite{bronnikov2018nonlinear}.

\textbf{Dynamical stability and field synergy.} Beyond static structural integrity, the physical viability of any black hole model relies on its dynamical stability against external linear perturbations. Regular black hole geometries coupled to NED are often susceptible to dynamical instabilities arising from anomalous couplings that generate negative wells in the effective potential. In our investigation of odd-parity (axial) perturbations, we demonstrated a consistent interaction between the background geometry and the perturbing field. The derived Regge-Wheeler-like effective potential remains positive and convex everywhere outside the event horizon for the entire physically allowed parameter space ($q_m \le 0.76$). Our numerical time-domain integration exhibited a stable quasi-normal ringing phase followed by a bounded exponential decay ($|\Psi| \sim e^{-\omega_I t}$). This numerical evidence was independently corroborated by the semi-analytical WKB approximation, which yielded negative imaginary components for the fundamental frequencies ($\text{Im}(\omega) < 0$). Thus, we conclude that the proposed configuration is globally dynamically stable \cite{bronnikov2024stability_epjc, toshmatov2018quasinormal, toshmatov2017stability}.

\textbf{Thermodynamic remnants and future perspectives.} Finally, our thermodynamic analysis revealed a strict mass gap ($M \ge M_{min}$) associated with the extremal configuration. This mandates that the semi-classical Hawking evaporation process must terminate as the black hole radiates its mass down to the extremal limit $M = M_{min}$, leaving behind a stable, massive thermodynamic remnant. The existence of such remnants provides a theoretical mechanism for resolving the black hole information loss paradox. Future investigations could extend this work by applying the Newman-Janis algorithm to construct the rotating (Kerr-like) generalization of this regular black hole, or by examining its response to even-parity (polar) gravitational perturbations, which would further solidify its astrophysical relevance.

\section*{Acknowledgments}

The authors express their sincere gratitude to Kirill Bronnikov for valuable comments and assistance in preparing this article.  The authors also express their gratitude to the Ministry of Higher Education, Science and Innovation of the Republic of Uzbekistan for the support under project No. FZ-20200929385.

\printbibliography
\end{document}